\documentclass{iucrjournals}
\pdfoutput=1 
\nolinenumbers
\newcommand{\rr}{\bm r}
\newcommand{\qq}{\bm q}
\newcommand{\kk}{\bm k}
\newcommand{\mvec}{\bm m}
\newcommand{\Mvec}{\bm M}
\newcommand{\Ms}{M_{\mathrm S}}

\newcommand{\dd}{\mathrm d}

\usepackage{amsmath,amssymb,bm}
\usepackage{graphicx}
\usepackage{hyperref}
\hypersetup{%
    pdfborder = {0 0 0}
}
\usepackage{xcolor}
\usepackage{tabularx,booktabs,array}
\usepackage{float}
\usepackage{subcaption}
\usepackage{caption}
\usepackage{microtype}

\graphicspath{{./figures/}}
 
\author[a]{V.D. Zaporozhets\IUCrCemaillink{vladdz@donfti.ru}\IUCrOrcidlink{xxxx-xxxx-xxxx-xxxx}}%
\author[a]{K.L. Metlov\IUCrEmaillink{metlov@donfti.ru}\IUCrOrcidlink{0000-0002-3929-5665}}%

\affil[a]{Galkin Donetsk Institute for Physics and Engineering, 72 R. Luxembourg str., Donetsk, Russia 83114}

\begin{document} 
\title{Magnetic small-angle neutron scattering by a nanocrystalline ferromagnet with anisotropic exchange interaction}
\maketitle

\begin{synopsis}
A micromagnetic framework of magnetic small-angle neutron scattering is developed that incorporates weak symmetric anisotropic exchange and predicts new angular harmonics in magnetic SANS cross sections of polycrystalline ferromagnets.
\end{synopsis}

\begin{abstract}
A micromagnetic framework for magnetic small-angle neutron scattering (SANS) is presented that accounts for weak symmetric anisotropic exchange in centrosymmetric nanocrystalline ferromagnets. The exchange interaction is expressed via a general fourth-rank tensor decomposed into isotropic and deviatoric parts. We start with the exchange energy and effective field, assuming weakly fluctuating in space saturation magnetization, solve micromagnetic problem to find spatial distribution of local magnetization vector and compute the averaged (over random orientations of nanocrystals) SANS cross sections. The isotropic part reproduces the classical Heisenberg-type SANS response, while non-zero deviatoric part of the exchange tensor gives rise to new angular harmonics in the magnetic SANS cross section. As a specific example, analytical response functions for an exchange tensor with hexagonal symmetry in perpendicular and parallel scattering geometries are derived. The results provide a basis for identifying and quantifying symmetric exchange anisotropy in magnetic SANS experiments.
\end{abstract}


\section{Introduction}

Magnetic small-angle neutron scattering (SANS) is an experimental technique for
characterizing spin textures, magnetic correlations, and magnetization
inhomogeneities in nanostructured ferromagnets and magnetic alloys.
By analyzing the angular dependence of the scattered intensity in reciprocal
space, magnetic SANS provides access to magnetization fluctuations on length
scales ranging from a few up to several hundred nanometers~\cite{Michels2021}.

Theoretical description of magnetic SANS is commonly based on
micromagnetic continuum theory~\cite{Michels2014}.
Within this framework, dominant sources of spin misalignment include
magnetostatic interactions, random magnetic anisotropy, exchange interaction,
and spatial variations of saturation magnetization.
Micromagnetic theory of magnetic SANS has been successfully applied to a wide range of bulk and
nanostructured ferromagnets~\cite{Michels2021}.
Several extensions of this theory have been proposed to account for additional
effects, such as macroscopic uniaxial anisotropy~\cite{Zaporozhets2022}, higher-order terms in the perturbative expansion of the
micromagnetic equations~\cite{MetlovMichels2015} and effects of the Dzyaloshinskii--Moriya interaction (DMI)~\cite{MichelsDMI2019}.

In all these approaches, however, the exchange interaction itself is treated
within the isotropic Heisenberg approximation, characterized by a single scalar
exchange stiffness and the corresponding exchange length.
This approximation considerably simplifies the theory and has proven adequate
for many applications.
Nevertheless, from a microscopic point of view, it is not always justified.

In real magnetic materials, the exchange interaction may acquire a directional dependence primarily through spin–orbit coupling, whose effect is enabled and modulated by reduced lattice symmetry, anisotropic chemical bonding, and local lattice distortions.
Such effects are commonly described in terms of anisotropic exchange
stiffness.
The influence of exchange anisotropy has been investigated in various contexts,
including spin-wave dispersion, domain-wall structure, and magnetic phase
stability~\cite{Belashchenko2004,Toga2018}.
In the context of magnetic SANS, anisotropic exchange is expected to modify the
angular dependence of the scattering cross section and to generate angular
harmonics that are absent in the isotropic theory.

At the continuum level, the most general exchange energy density can be written
as the sum of two distinct contributions~\cite{akhiezer1968spinwaves}.
The first is an antisymmetric term that is linear in spatial gradients and gives
rise to the DMI, which is allowed only in
non-centrosymmetric crystals.
The second is a symmetric term that is quadratic in gradients and is described
by a fourth-rank exchange tensor.
Although the influence of DMI on magnetic SANS has been discussed in several recent
studies~\cite{MichelsDMI2019, Quan_2020}, the role of symmetric anisotropic exchange has received comparatively
little attention.
In centrosymmetric ferromagnets, where DMI is forbidden by symmetry, symmetric
exchange anisotropy represents the leading correction to the isotropic
Heisenberg interaction.

The aim of the present work is to develop a micromagnetic theory of magnetic
SANS that consistently accounts for weak symmetric anisotropic exchange in
polycrystalline ferromagnets.
Starting from a general fourth-rank exchange tensor, we separate the isotropic
and deviatoric contributions and derive a compact expression for the effective
exchange field that remains valid in the presence of spatially inhomogeneous
material parameters.
Within a linearized micromagnetic framework, the isotropic part of the exchange
tensor reproduces the standard Heisenberg-type SANS response, while all
the anisotropic corrections originate from the deviator part of the exchange
tensor.

We show that symmetric exchange anisotropy leads to additional angular
harmonics in the magnetic SANS cross section, whose structure depends on the
symmetry of the exchange tensor and on the scattering geometry.
As a specific example, we analyze the case of hexagonal exchange symmetry and
derive explicit analytical expressions for the corresponding anisotropic
response functions in both perpendicular and parallel scattering geometries.
The obtained results provide a framework for identifying and
quantifying the effects of weak symmetric exchange anisotropy in magnetic SANS
experiments.

\section{Magnetic small-angle neutron scattering}

For an ordered ferromagnet well below the Curie temperature, the local magnetization satisfies $|\Mvec(\rr)|=\Ms(\rr)$ and can be written as $\Mvec(\rr)=\Ms(\rr)\,\mvec(\rr)$ with $|\mvec|=1$.
In a perfectly uniform medium, $\mvec$ aligns with $\bm H$.
In real nanostructured materials, defects and spatial variations in material parameters produce spin misalignment, which causes magnetic SANS at finite $q$.

In a typical magnetic SANS experiment~\cite{Michels2014}, neutrons with incident wave vector $\kk$ pass through a sample and are scattered in the direction of an outgoing wave vector $\kk'$.
The momentum-transfer vector is
\begin{equation}
\qq = \kk' - \kk .
\label{eq:qdef}
\end{equation}
The scattering vector $\qq$ selects spatial Fourier components of the magnetization distribution that contribute to the measured intensity.

The total macroscopic (unpolarized) SANS cross section can be decomposed into
\begin{equation}
\frac{\dd\Sigma}{\dd\Omega}
=
\left(\frac{\dd\Sigma}{\dd\Omega}\right)_{\!\mathrm{res}}
+
\left(\frac{\dd\Sigma}{\dd\Omega}\right)_{\!\mathrm{M}},
\label{eq:Sigma_total}
\end{equation}
where $(\dd\Sigma/\dd\Omega)_{\mathrm{res}}$ denotes a field-independent residual (nuclear and magnetic)
background, and $(\dd\Sigma/\dd\Omega)_{\mathrm{M}}$ is the magnetic
spin-misalignment contribution that vanishes in the high-field limit.
\begin{figure}[ht] %
\label{fig:figure1}
\begin{center}
\includegraphics[width=0.5\textwidth]{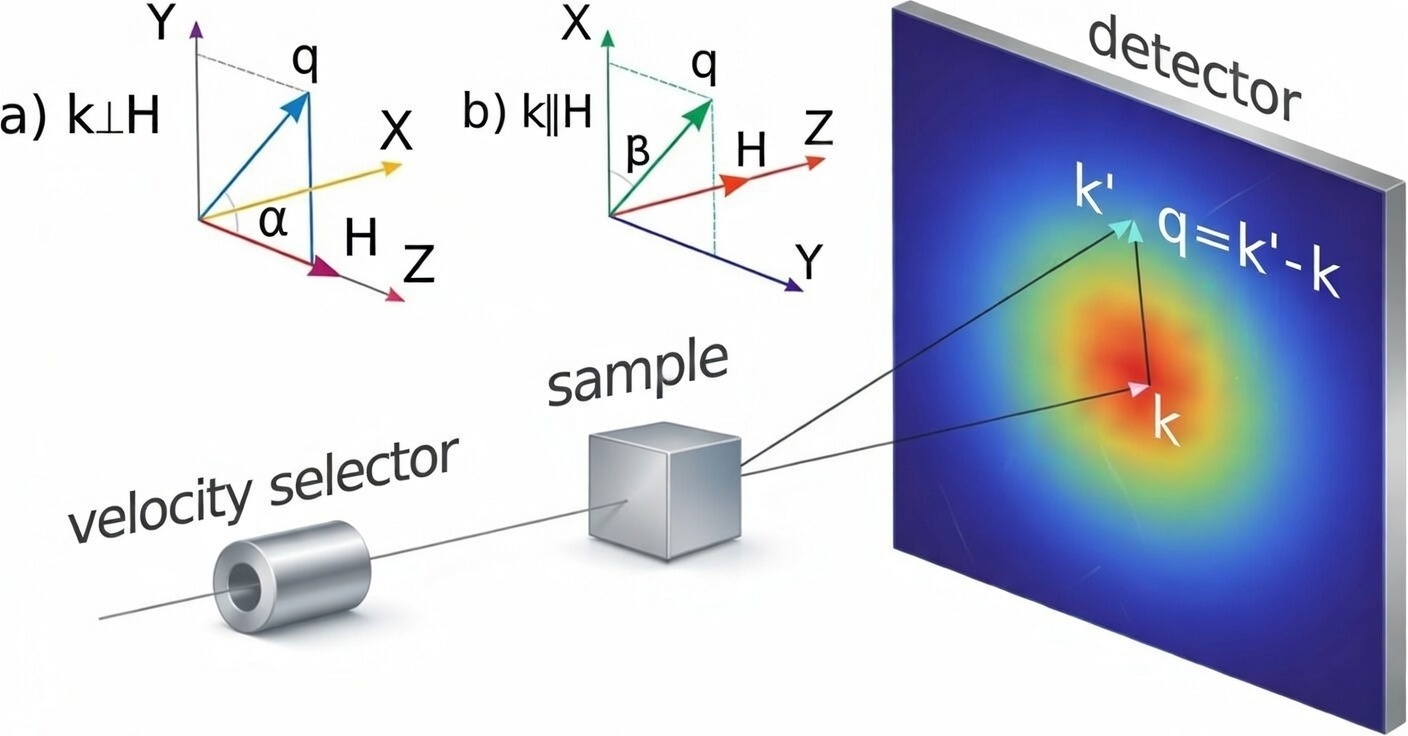} 
\end{center}
\caption{Typical geometry of a magnetic SANS experiment.
Neutrons with incident wave vector $\mathbf{k}$ are scattered into the
direction $\mathbf{k}'$, giving rise to the scattering vector
$\mathbf{q}=\mathbf{k}'-\mathbf{k}$.
The applied magnetic field $\mathbf{H}$ can be oriented either perpendicular
or parallel to the beam direction.
In the perpendicular geometry, $\mathbf{H}$ is directed along the
$Z$ axis, and the detector plane coincides with the $Y$--$Z$ plane.} 
\label{fig:magnetic_sans_geometry}
\end{figure}

Figure~\ref{fig:magnetic_sans_geometry} shows that magnetic SANS experiments are commonly performed in two scattering geometries:
(i) the perpendicular geometry, where the incident neutron beam is oriented
perpendicular to the applied magnetic field, $\mathbf k \perp \mathbf H$, and
(ii) the parallel geometry, where $\mathbf k \parallel \mathbf H$.
In both cases, a Cartesian laboratory coordinate system is introduced such that
the scattering vector $\mathbf q$ lies in the detector plane.

The orientation of $\mathbf q$ in the detector plane is conveniently described
by its magnitude $q=|\mathbf q|$ and an azimuthal angle $\alpha$ or $\beta$, measured with
respect to the $X$ axis.
Consequently, the components of the dimensionless scattering vector
$\mathbf q/q$ are expressed as
\[
(x_q, y_q, z_q) =
\begin{cases}
(0, \cos\alpha, \sin\alpha), & \text{perpendicular geometry}, \\[4pt]
(\cos\beta, \sin\beta, 0), & \text{parallel geometry}.
\end{cases}
\]

In perpendicular geometry, the unpolarized magnetic spin-misalignment can be written in the following form~\cite{Michels2014}
\begin{equation}
\frac{\dd\Sigma^{\perp}_{\mathrm M}}{\dd\Omega}
=
8\pi^3V\, b_{\mathrm H}^2
\Big(
|\widetilde M_X|^2
+
|\widetilde M_Y|^2 \cos^2\alpha
+
\big(|\widetilde M_Z|^2-|\widetilde M_{\mathrm S}|^2\big)\sin^2\alpha
-
\Re\{\widetilde M_Y \widetilde M_Z^\ast\}\sin 2\alpha
\Big),
\label{eq:Sigma_perp}
\end{equation}
where $V$ is the scattering volume, $b_{\mathrm H}$ is the magnetic scattering
length per Bohr magneton, and $\widetilde M_{X,Y,Z}(\mathbf q)$ are Fourier components of
the magnetization. In Eq.~\eqref{eq:Sigma_perp}, $\Re\{\cdots\}$ denotes the real part and the
superscript $\ast$ indicates the complex conjugation.

In the parallel geometry, the corresponding unpolarized magnetic spin-misalignment cross
section reads
\begin{equation}
\frac{\dd\Sigma^{\parallel}_{\mathrm M}}{\dd\Omega}
=
8\pi^3V\, b_{\mathrm H}^2
\Big(
|\widetilde M_X|^2 \sin^2\beta
+
|\widetilde M_Y|^2 \cos^2\beta
+
\big(|\widetilde M_Z|^2-|\widetilde M_{\mathrm S}|^2\big)
-
\Re\{\widetilde M_X \widetilde M_Y^\ast\}\sin 2\beta
\Big).
\label{eq:Sigma_parallel}
\end{equation}
Magnetic small-angle neutron scattering experiments can be performed in
various scattering geometries depending on the relative orientation of the
applied magnetic field, the incident neutron beam, and the detector plane.
In the present work, however, we restrict ourselves to the two experimentally
relevant geometries mentioned above.

The Fourier transform convention used throughout this work is
\begin{equation}
\widetilde F(\mathbf q)
=
\frac{1}{V}\int_V F(\mathbf r)\,e^{-\mathrm i \mathbf q\cdot \mathbf r}\,\dd^3 r,
\qquad
F(\mathbf r)=\sum_{\mathbf q}\widetilde F(\mathbf q)\,e^{\mathrm i \mathbf q\cdot \mathbf r}.
\label{eq:FT}
\end{equation}
Equations~\eqref{eq:Sigma_perp}--\eqref{eq:FT} show that the central theoretical
task is to compute the Fourier components
$\widetilde M_X$, $\widetilde M_Y$, and $\widetilde M_Z$
from a micromagnetic model. 
In the following sections, we formulate the exchange energy in a general
tensorial form and derive the corresponding effective field and its Fourier
representation, which provides the starting point for the linearized Brown
equations and for the analytical calculation of magnetic SANS response
functions in both scattering geometries.

\section{Symmetric anisotropic exchange energy}

In a centrosymmetric ferromagnet, the most general continuum exchange energy density that is quadratic in spatial derivatives of the magnetization can be written in tensorial form~\cite{akhiezer1968spinwaves} as
\begin{equation}
e_{\mathrm{EX}}
=
\frac{1}{2}
\sum_{\alpha\beta\gamma\delta}
C_{\alpha\beta\gamma\delta}
\,
\partial_\beta m_\alpha
\,
\partial_\delta m_\gamma ,
\label{eq:ex_energy_density}
\end{equation}
where Greek indices denote Cartesian components, $\partial_\alpha \equiv \partial / \partial x_\alpha$, and $C_{\alpha\beta\gamma\delta}$ is a fourth-rank exchange tensor.
The tensor is symmetric with respect to the index pairs $(\alpha,\beta)$ and $(\gamma,\delta)$ and reflects the symmetry of the crystal point group. This approach unifies various possible symmetries within a single framework. The components $C_{\alpha\beta\gamma\delta}$ quantitatively determine how the change in the magnetization component $m_\alpha$ along the $x_\beta$ direction is related to the change in $m_\gamma$ along $x_\delta$.
For an isotropic ferromagnet, the tensor reduces to
\begin{equation}
C_{\alpha\beta\gamma\delta}
=
C \, \delta_{\alpha\gamma}\delta_{\beta\delta},
\end{equation}
where $C$ is the conventional scalar exchange stiffness.
In the general case, $C_{\alpha\beta\gamma\delta}$ contains both isotropic and anisotropic contributions.

In nanostructured ferromagnets, saturation magnetization $\Ms(\rr)$ and exchange constant $C(\rr)$ may vary spatially due to compositional fluctuations or multiphase microstructure. We introduce~\cite{Michels2021}
\begin{equation}
M_S(\mathbf r)=M_0\left(1+I_M(\mathbf r)\right), \qquad |I_M|\ll 1,
\label{eq: SaturationMagnetization}
\end{equation}
where $I_M$ describes its weak inhomogeneities and is zero on average $\langle I_M\rangle = 0$. Using a derivative operator (see the Appendix~\ref{app:derivative} for definition) that takes into account the inhomogeneity of $M_S$:
\begin{equation}
\mathcal D_\beta M_\alpha=
\partial_\beta M_\alpha
- M_\alpha\,\partial_\beta\ln\frac{M_S}{M_0},
\end{equation}
makes it possible to compactly write down the exchange energy as
\begin{equation}
e_{\mathrm{EX}}(\mathbf r)
=
\frac{\gamma_B}{2M_S^2}\sum_{\alpha,\beta,\gamma,\delta}
C_{\alpha\beta\gamma\delta}(\mathbf r)
(\mathcal D_\beta M_\alpha)(\mathcal D_\delta M_\gamma).
\end{equation}
This form is particularly useful for deriving the effective exchange field via functional differentiation.

\section{Effective exchange field}

From the effective field definition, calculating all the derivatives and using the symmetry of the tensor $C_{\alpha\beta\gamma\delta}$, we obtain (see the Appendix~\ref{app:exchfield} for details) a compact expression
\begin{equation}
H_{\mathrm{EX},\alpha}
=
\frac{\gamma_B}{\mu_0}
\left[
\partial_\beta T_{\alpha\beta}
+
(\partial_\beta\ln M_S)\,T_{\alpha\beta}
\right],
\end{equation}
where the quantity 
\begin{equation}
    T_{\alpha\beta}
= \frac{C_{\alpha\beta\gamma\delta}}{M_S^2}\,
\mathcal D_\delta M_\gamma
\label{eq: AalphaBeta}
\end{equation}
is conceptually similar to a stress tensor in elasticity theory. The factor $\gamma_B$ is introduced~\cite{AharoniBook} to conveniently account for different possible systems of magnetic units.
For example, in the SI system $\mu_0$ denotes the permeability of vacuum and
$\gamma_B = 1$, whereas in the CGS system $\mu_0 = 1$ and $\gamma_B = 4\pi$.

Having obtained the general formula for the effective exchange field, we proceed to calculate its Fourier transform, which is necessary for further derivation of the angular dependences of the magnetic SANS cross sections. The key step here is to expand the quantity $T_{\alpha\beta}$ in terms of smallness of $I_M(\mathbf r)$ and isolate only those terms that contribute in the first order of perturbation theory.

Expanding the expression $1/M_S^2$ up to the first order in saturation magnetization fluctuations
\begin{equation}
\frac{1}{M_S^2}
=
\frac{1}{M_0^2} (1 - 2 I_M)
=
\frac{1}{M_0^2} + O(I_M),
\end{equation}
and substituting it in~\eqref{eq: AalphaBeta} together with~\eqref{eq: SaturationMagnetization}
inside the functional derivative $\mathcal D_\beta$, we obtain:
\begin{equation}
T_{\alpha\beta} =
\frac{C_{\alpha\beta\gamma\delta}}{M_s^2}
\, \mathcal D_\delta M_\gamma
=
\frac{C_{\alpha\beta\gamma\delta}}{M_0^2}
\left( \partial_\delta M_\gamma - M_\gamma \partial_\delta I_M \right)
+ O(I_M^2).
\end{equation}

Now, let us treat the transversal and longitudinal components separately.
For $\gamma = Z$:
\begin{equation}
\partial_\delta M_Z = M_0 \partial_\delta I_M,
\quad
\Rightarrow
\quad
\mathcal D_\delta M_Z
= M_0 \partial_\delta I_M - M_0 (1+I_M)\partial_\delta I_M
= O(I_M^2).
\end{equation}
That is, the longitudinal part completely disappears in the first order.

For $\gamma = X, Y$:
\begin{equation}
\mathcal D_\delta M_{X,Y}
= \partial_\delta M_{X,Y}^{(1)}
 - M_{X,Y}^{(1)}\partial_\delta I_M,
\end{equation}
where $M_{X,Y}^{(1)}$ are magnetization components that are of the same order as $I_M$. Since the second term contains the product of two small quantities, it is of the second order and is discarded:
\begin{equation}
\mathcal D_\delta M_{X,Y} \approx \partial_\delta M_{X,Y}^{(1)}
\end{equation}

Taking these expressions into account, we can write the following formula for $T_{\alpha\beta}$
\begin{equation}
T_{\alpha\beta}
=
\frac{C_{\alpha\beta\gamma\delta}}{M_0^2} \, \partial_\delta M_\gamma.
\end{equation}
Thus, the expression for the effective exchange field in the first order takes the form
\begin{equation}
H_{\mathrm{EX},\alpha}
=
\frac{1}{\mu_0}
\partial_\beta T_{\alpha\beta}
=
\frac{1}{\mu_0 M_0^2}
C_{\alpha\beta\gamma\delta}\,
\partial_\beta \partial_\delta M_\gamma.
\end{equation}
To get the expressions for transverse magnetization components, one needs to compute the Fourier image of the above effective field. Using $\partial_\beta \to i q_\beta$, we have
\begin{equation}
\widetilde{H}_{\mathrm{EX},\alpha}
=
\frac{1}{\mu_0 M_0^2}
C_{\alpha\beta\gamma\delta}
(i q_\beta)(i q_\delta)
\widetilde{M}
=
-\frac{1}{\mu_0 M_0^2}
C_{\alpha\beta\gamma\delta}\,
q_\beta q_\delta\,
\widetilde{M}.
\end{equation}
Let us introduce a matrix $K_{\alpha\gamma}$
\begin{equation}
K_{\alpha\gamma}(\bm{q})
=
\frac{1}{\mu_0 M_0^2}
C_{\alpha\beta\gamma\delta}\, q_\beta q_\delta,
\end{equation}
to represent the effective field in a compact form:
\begin{equation}
\widetilde{H}_{\mathrm{EX},\alpha}
=
-K_{\alpha \gamma}(\bm{q})
\widetilde{M}_{\gamma}.
\end{equation}
This representation is especially convenient, since all the exchange
contributions enter the linearized Brown's equations exclusively through the
matrix $K_{\alpha\gamma}(\mathbf q)$. As a result, the influence of the 
anisotropic exchange on the magnetic response is entirely governed by the
tensorial structure of $K(\mathbf q)$.
Because the exchange tensor
$C_{\alpha\beta\gamma\delta}$ is symmetric with respect to the permutations of
index pairs, $K_{\alpha\gamma}(\mathbf q)$ is a real symmetric matrix for each fixed
$\mathbf q$.

\section{Deviatoric part of the exchange tensor}
Now we decompose the tensor $C_{\alpha\beta\gamma\delta}$ into the diagonal part and the deviator:
\begin{equation}
C_{\alpha\beta\gamma\delta}
=
C_0\, \delta_{\alpha\gamma}\delta_{\beta\delta}
+
\xi\, D_{\alpha\beta\gamma\delta},
\end{equation}
where $\xi$ is a small parameter,
and $\mathrm{Tr}(D)=0$.
The contribution of the isotropic part to the $K$ function gives
\begin{equation}
K^{(0)}_{\alpha\gamma}(q)
=
\frac{1}{\mu_0 M_0^2}\,
C_0\, q^2\, \delta_{\alpha\gamma}
=
\gamma_B L_0^2\, q^2\, \delta_{\alpha\gamma},
\end{equation}
where $L_0=\sqrt{2C/(\gamma_B\mu_0M_0^2)}$ is the exchange length. The anisotropic contribution has the form
\begin{equation}
\label{eq:deviatoricKernel}
\Delta K_{\alpha\gamma}(q)
=
\frac{1}{\gamma_B \mu_0 M_0^2}
D_{\alpha\beta\gamma\delta}
q_\beta q_\delta.
\end{equation}

These expressions are currently written in a coordinate system associated with the crystallite. However, in a polycrystalline ferromagnet, each grain has a random orientation of its crystallographic axes, which, as a rule, do not coincide with the orientation of the laboratory frame, defined by the direction of the external magnetic field and the neutron beam.

After the isotropic and deviatoric parts of the exchange tensor are separated, it becomes possible to write down the effective field in the laboratory coordinate system and clearly trace which combinations of tensor components and fluctuations of magnetic quantities determine the angular dependence of the magnetic cross sections of SANS.
Let the rotation of an individual grain relative to the laboratory system be given by the Euler matrix $\mathcal R(\phi,\theta,\psi)$. Then the transition from crystallographic indices $(\alpha\beta\gamma\delta)$ to laboratory indices $(ijkl)$ is carried out according to the standard rule:
\begin{equation}
C^{(R)}_{ijkl}
=
\mathcal R_{i\alpha}
 \mathcal R_{j\beta}
\mathcal R_{k\gamma}
\mathcal R_{l\delta}
\,
C_{\alpha\beta\gamma\delta}.
\end{equation}
Substituting this expression into the isotropic part of the function $K_{\alpha\gamma}(q)$, we obtain:
\begin{equation}
K^{(0)}_{ij}(q)
=
\mathcal R_{i\alpha} \mathcal R_{j\gamma} K^{(0)}_{\alpha\gamma}(q)
=
C_0 q^2 \mathcal R_{i\alpha} \mathcal R_{j\gamma}\delta_{\alpha\gamma}
=
C_0 q^2 \delta_{ij}.
\end{equation}
Thus, the isotropic contribution takes the same form
in any coordinate system. Physically, this means that ordinary (scalar) exchange is independent of the crystal lattice orientation.
For the deviatoric part, we have:
\begin{equation}
    \Delta K_{ij}(q)
= R_{i\alpha}
\mathcal R_{j\beta}
\mathcal R_{k\gamma}
\mathcal R_{l\delta}
\,
D_{\alpha\beta\gamma\delta} q_k q_l
\end{equation}
where the indices for $\mathbf{q}$ are changed to the laboratory frame because only the tensor $D_{\alpha\beta\gamma\delta}$ must be rotated, whereas $\mathbf{q}$ remains fixed. This way, the expression for $K$ becomes
\begin{equation}
\Delta K_{ij}(q)
=
\mathcal R_{i\alpha} \mathcal R_{j\gamma}
\,
\Delta K_{\alpha\gamma}(q),
\end{equation}
and the effective exchange field assumes its final form:
\begin{equation}
\widetilde H_{\mathrm{EX},i}
=
-
\gamma_B
\Big[
K^{(0)}_{ij}
+
\xi\, \Delta K_{ij}
\Big]
\widetilde M^{(1)}_j(q),
\end{equation}
where $\Delta K_{ij}$ includes all the effects of symmetric exchange
in the first order of the small parameter~$\xi$.
The components of the resulting expression can be written explicitly:
\begin{equation}
\widetilde H_{\mathrm{EX},i}
=
-
\gamma_B\Big[
 L_0^2 q^2 \widetilde M_i^{(1)}
+
\xi
\left(
\Delta K_{iX} \widetilde M_X^{(1)}
+
\Delta K_{iY} \widetilde M_Y^{(1)}
+
\Delta K_{iZ} \widetilde I_M
\right)
\Big],
\end{equation}
where $\widetilde I_M$ is the Fourier transform of the inhomogeneity function. These expressions show that the appearance of exchange anisotropy leads to new types of connection between transverse and longitudinal magnetization fluctuations. Terms of the form
\begin{equation}
\Delta K_{XZ}\,\widetilde I_M,
\qquad
\Delta K_{ZX}\,\widetilde M_X^{(1)},
\qquad
\Delta K_{YZ}\,\widetilde M_Y^{(1)},
\end{equation}
form new angular dependencies in magnetic SANS cross sections.

Thus, the exchange tensor deviator plays the role of
a source of anisotropy in magnetic SANS,
and the observed scattering pattern reflects the
averaged over all orientations of crystallites effect of these corrections.

\section{Brown's equations}

Substituting the expressions obtained for the effective exchange field,
together with the contributions of the remaining micromagnetic interactions~\cite{MetlovMichels2015}, into the Brown equations and solving them
to first order in the small parameters, we obtain the following expressions
for the magnetization components in Fourier space:
\begin{align}
\widetilde m_{X}^{(1)}
&=
\frac{
\bigl(\widetilde A_X - \widetilde{I}_M x_q y_q + \widetilde{I}_M \Delta K_{XZ}\,\xi\bigr)\,G_Y
+
X_q\,
\bigl(\widetilde{I}_M x_q z_q - \widetilde{I}_M \Delta K_{YZ}\,\xi - \widetilde A_Y\bigr)
}{
(h_q + x_q^2 - \Delta K_{XX}\xi)(h_q + y_q^2 - \Delta K_{YY}\xi)
- (x_q y_q - \Delta K_{XY}\xi)^2
}\label{eq:mX1},
\\[6pt]
\widetilde m_{Y}^{(1)}
&=
\frac{
\bigl(\widetilde A_Y - \widetilde{I}_M x_q y_q + \widetilde{I}_M \Delta K_{YZ}\,\xi\bigr)\,G_X
+
X_q\,
\bigl(\widetilde{I}_M y_q z_q - \widetilde{I}_M \Delta K_{XZ}\,\xi - \widetilde A_X\bigr)
}{
(h_q + y_q^2 - \Delta K_{YY}\xi)(h_q + x_q^2 - \Delta K_{XX}\xi)
- (x_q y_q - \Delta K_{XY}\xi)^2
}\label{eq:mY1}.
\end{align}
Here $h_q = h + L_0^2 q^2$, and the contribution of random magnetic anisotropy is represented by the quantities
$\widetilde A_{X,Y} = u_{X,Y}\widetilde Q_r$, where $\widetilde Q_r$ is the
Fourier transform of the random anisotropy quality factor, which is assumed
to be of the same order of smallness as $I_M$.
For the case of random uniaxial anisotropy, the coefficients are given by
$u_{X,Y}=d_{X,Y} d_Z$, where $d_{X,Y,Z}(\mathbf r)$ are unit vectors along the
local anisotropy axes.

For compactness, we have introduced the auxiliary quantities
\begin{equation}
G_X = h_q + x_q^{2} - \Delta K_{XX}\,\xi,
\qquad
G_Y = h_q + y_q^{2} - \Delta K_{YY}\,\xi,
\end{equation}
and
\begin{equation}
X_q = x_q y_q - \Delta K_{XY}\,\xi.
\end{equation}

The expressions above for $\widetilde m_X^{(1)}$ and $\widetilde m_Y^{(1)}$
constitute the central result required for the calculation of magnetic
small-angle neutron scattering cross sections.
In order to obtain observable macroscopic quantities in a polycrystalline sample, these expressions must
be averaged over the crystallite orientations.
This requires averages of various tensor and vector products, such as
$\langle A_{X/Y} \Delta K_{ij} \rangle$ and $\langle \Delta K_{ij} \rangle$. They depend explicitly on the symmetry of the exchange tensor, or, more
precisely, on the symmetry of its deviatoric part and on the chosen scattering geometry.
Owing to the smallness of the anisotropy parameter $\xi$, this averaging
procedure can be carried out perturbatively.

Substituting the expressions for the transverse components of the
magnetization~\eqref{eq:mX1} and~\eqref{eq:mY1} into the formula for the perpendicular
magnetic SANS cross section and expanding in powers of $\xi$, the 
cross section can be split into two parts
\begin{equation}
\frac{d\Sigma^{M}_{\perp}}{d\Omega} 
=
\left.
\frac{d\Sigma^{M}_{\perp}}{d\Omega}
\right|_{HEIS}
+
\left.
\frac{d\Sigma^{M}_{\perp}}{d\Omega}
\right|_{A},
\end{equation}
where the first term corresponds to the well-known Heisenberg-type
exchange contribution~\cite{Michels2021}, while the second term represents the anisotropic
corrections.

The anisotropic contribution can be further expanded into a series over the small
parameter $\xi$,
\begin{equation}
\left.
\frac{d\Sigma^{M}_{\perp}}{d\Omega}
\right|_{A}
=
\left.
\frac{d\Sigma^{M}_{\perp}}{d\Omega}
\right|_{\xi}
+
\left.
\frac{d\Sigma^{M}_{\perp}}{d\Omega}
\right|_{\xi \kappa}
+
\left.
\frac{d\Sigma^{M}_{\perp}}{d\Omega}
\right|_{\xi^2}
+ \ldots .
\end{equation}
The leading anisotropic contribution, linear in $\xi$, takes the form
\begin{equation}
\left.
\frac{d\Sigma^{M}_{\perp}}{d\Omega}
\right|_{\xi}
=
\frac{8\pi^3}{V}\, b_{\mathrm H}^2 M_0^2 \widetilde{I}_M^{\,2} \, \xi\,
\frac{
2 y_q z_q (h_q +1)
\big(
K_{YY} y_q z_q - K_{YZ} (h_q + y_q^{2})
\big)
}{
(h_q + y_q^{2})^{3}
}.
\end{equation}
The mixed contribution proportional to $\kappa \xi$ can be written in a compact
form as a scalar product of two vectors,
$\mathbf u = (u_X,u_Y)$ and $\boldsymbol{\Phi} = (\Phi_X,\Phi_Y)$,
\begin{equation}
\left.
\frac{d\Sigma^{M}_{\perp}}{d\Omega}
\right|_{\kappa\xi}
=
\frac{8\pi^3}{V}\, b_{\mathrm H}^2 M_0^2 \widetilde{I}_M^{\,2}
\, \kappa \xi \,
(\mathbf u \cdot \boldsymbol{\Phi}),
\label{eq:Sigma_kappa_xi}
\end{equation}
where the components of the vector $\boldsymbol{\Phi}$ are given by
\begin{align}
\Phi_X
&=
-
\frac{2 K_{YY}\, y_q\, z_q}{\left(h_q + y_q^{2}\right)^2}
+
\frac{2 h_q K_{YZ}\, z_q^{2}}{\left(h_q + y_q^{2}\right)^3}
+
\frac{2 K_{YY}\, y_q^{2}\, z_q^{2}}{\left(h_q + y_q^{2}\right)^3}
-
\frac{4 K_{YY}\, y_q\, z_q^{3}}{\left(h_q + y_q^{2}\right)^3},
\\[6pt]
\Phi_Y
&=
\frac{2 K_{XZ}}{h_q^{2}}
-
\frac{2 K_{XY}\, y_q\, z_q}{\left(h_q + y_q^{2}\right)^2}
-
\frac{2 K_{XY}\, y_q\, z_q}{h_q \left(h_q + y_q^{2}\right)^2}
-
\frac{2 K_{XY}\, y_q^{3}\, z_q}{h_q^{2} \left(h_q + y_q^{2}\right)}
\nonumber\\
&\quad
-
\frac{2 K_{XY}\, y_q\, z_q^{3}}{\left(h_q + y_q^{2}\right)^3}
-
\frac{2 K_{XY}\, y_q^{3}\, z_q^{3}}{h_q \left(h_q + y_q^{2}\right)^3}.
\end{align}

The full expression for the contribution quadratic in $\xi$ is lengthy and is
therefore relegated to Appendix~\ref{app:chi2}.

It is important to emphasize that the parameters \(\xi\) and \(\kappa\) play fundamentally different roles in the present theory and are therefore treated differently.
The parameter \(\xi\) measures the relative strength of the symmetric anisotropic part of the exchange tensor with respect to its isotropic component.
Since symmetric exchange anisotropy originates from spin--orbit coupling, it is generally much weaker than the dominant Heisenberg exchange, which justifies a systematic expansion in the powers of \(\xi\).

In contrast, the parameter \(\kappa\) relates the random magnetic anisotropy quality factor to the general inhomogeneity function \(I_M(\mathbf r)\) that describes spatial variations of magnetic material parameters.
It does not represent an independent small interaction, but rather characterizes how random anisotropy couples to existing magnetization inhomogeneities.
As a result, \(\kappa\) is treated on the same footing as other sources of spin misalignment present in isotropic micromagnetic theory and is not used as an expansion parameter.
The terms proportional to \(\xi\kappa\) therefore describe the leading coupling between weak symmetric exchange anisotropy and random-anisotropy-induced spin disorder.

In order to obtain explicit angular dependencies, it is necessary to specify the symmetry of the tensor \(C_{\alpha\beta\gamma\delta}\) and to perform the corresponding averaging of the expressions entering
\(
\left.\mathrm d\Sigma^{\mathrm M}/\mathrm d\Omega\right|_{A}.
\)
In the present work, we investigate the influence of hexagonal exchange on the magnetic SANS cross sections.

\section{Hexagonal symmetry of the exchange tensor}

The number of independent exchange parameters is determined by the crystallographic point symmetry acting on the fourth-rank exchange tensor \(C_{\alpha\beta\gamma\delta}\).
The higher the crystal symmetry, the smaller the number of independent coefficients; conversely, a lower symmetry allows for a larger number of distinct exchange stiffnesses.
In the present section, we restrict ourselves to the conventional minimal representation of hexagonal symmetry, which captures the dominant anisotropic effects while preserving analytical transparency of the model.

In the crystallite coordinate system \((X',Y',Z')\), the hexagonal axis is directed along \(Z'\).
In this coordinate system, the deviatoric part of the hexagonal symmetry exchange tensor takes the form:
\begin{equation}
D_{\alpha\beta\gamma\delta}
=
\delta_{\alpha\gamma}\,
\Delta C
\left(
n_\beta n_\delta
-
\frac{1}{3}\delta_{\beta\delta}
\right),
\end{equation}
where \(\Delta C = C_Z - C_{\perp}\) is the difference between the longitudinal and transverse exchange constants, and \(n_\alpha=\delta_{\alpha Z}\) are the components of the unit vector
\(
\mathbf n=(0,0,1)
\)
along the hexagonal axis.

Substituting this expression into the definition of the deviatoric part of the kernel~\eqref{eq:deviatoricKernel}, one obtains for the hexagonal symmetry
\begin{equation}
\Delta K_{\alpha\gamma}
=
\delta_{\alpha\gamma}\,
\Delta C
\left[
(\mathbf n \cdot \mathbf q)^2
-
\frac{1}{3}q^2
\right],
\end{equation}
where the dot denotes the scalar product.

In order to obtain the expression for \(\Delta K_{\alpha\gamma}\) in the laboratory coordinate system, the tensor \(D_{\alpha\beta\gamma\delta}\) is transformed using the rotation matrix \(\mathcal{R}\),
\begin{equation}
D^{(R)}_{ijkl}
=
\mathcal{R}_{i\alpha}
\mathcal{R}_{j\beta}
\mathcal{R}_{k\gamma}
\mathcal{R}_{l\delta}
D_{\alpha\beta\gamma\delta}
=
\Delta C\,
\mathcal{R}_{i\alpha}
\mathcal{R}_{j\beta}
\mathcal{R}_{k\gamma}
\mathcal{R}_{l\delta}
\delta_{\alpha\gamma}
\left(
n_\beta n_\delta
-
\frac{1}{3}\delta_{\beta\delta}
\right).
\end{equation}

Using the orthogonality relations
\(
\mathcal{R}_{i\alpha}\mathcal{R}_{j\alpha}=\delta_{ij}
\)
and
\(
\hat{R}_{k\beta}\mathcal{R}_{l\beta}=\delta_{kl},
\)
one obtains
\begin{equation}
D^{(R)}_{ijkl}
=
\Delta C\,
\delta_{ij}
\left(
n_k n_l
-
\frac{1}{3}\delta_{kl}
\right),
\end{equation}
which has the same structure as in the crystallite coordinate system, but now with
\(
\mathbf n = (n_X,n_Y,n_Z)
\) expressed through laboratory coordinates.

Substituting the expression for \(D^{(R)}_{ijkl}\) into the formula for \(\Delta K_{ik}\), we get the following expression for the deviatoric part of the kernel \(K\) in the laboratory coordinate system:
\begin{equation}
\Delta K_{ik}
=
\Lambda_q^{\,2}
\left[
(\mathbf n \cdot \mathbf q)^2
-
\frac{1}{3} q^2
\right]
n_i n_k ,
\end{equation}
where
\begin{equation}
\Lambda_q^{\,2}
=
\frac{\Delta C}{\mu_0 \gamma_B M_0^{\,2}} .
\end{equation}
To obtain expressions for perpendicular cross sections, one needs to substitute expressions for $\Delta K^{(R)}_{ij}$ into~\eqref{eq:Sigma_perp} and carry out averaging procedure.

\subsection{Magnetic SANS in the presence of hexagonal exchange}

The averaging procedure required to obtain macroscopic magnetic SANS cross
sections consists of two steps: averaging over realizations of structural
inhomogeneities~\cite{MetlovMichels2015}, and averaging
over the orientations of crystallites in a polycrystalline sample.
The latter reduces to evaluating the averages of combinations such as
$\langle \Delta K^{(R)}_{XY} u_X(\mathbf q) \rangle$.
In the following, we assume that the directions of the hexagonal exchange axis
and the uniaxial anisotropy axis coincide.
Expressing the unit vectors $\bm n$ and $\bm d$ through the same set of
spherical angles and performing the corresponding angular integrations yields
the averaged scattering cross sections.

Following the formulation introduced in~\cite{Michels2014}, the
anisotropic part of the perpendicular magnetic SANS cross section can be
written as
\begin{equation}
\left. \frac{d\Sigma^{M}_{\perp}}{d\Omega} \right|_{A}
=
S_\xi R_{\xi}
+
S_{\xi \kappa} R_{\xi \kappa}
+
S_{\xi^2} R_{\xi^2}
+ \ldots ,
\end{equation}
where $R(\bm q,h_q)$ are referred to as response functions and $S(q)$ are the
corresponding scattering functions.

\begin{figure*}[h!]
\centering

\begin{subfigure}[t]{0.48\textwidth}
  \centering
  \includegraphics[width=\linewidth]{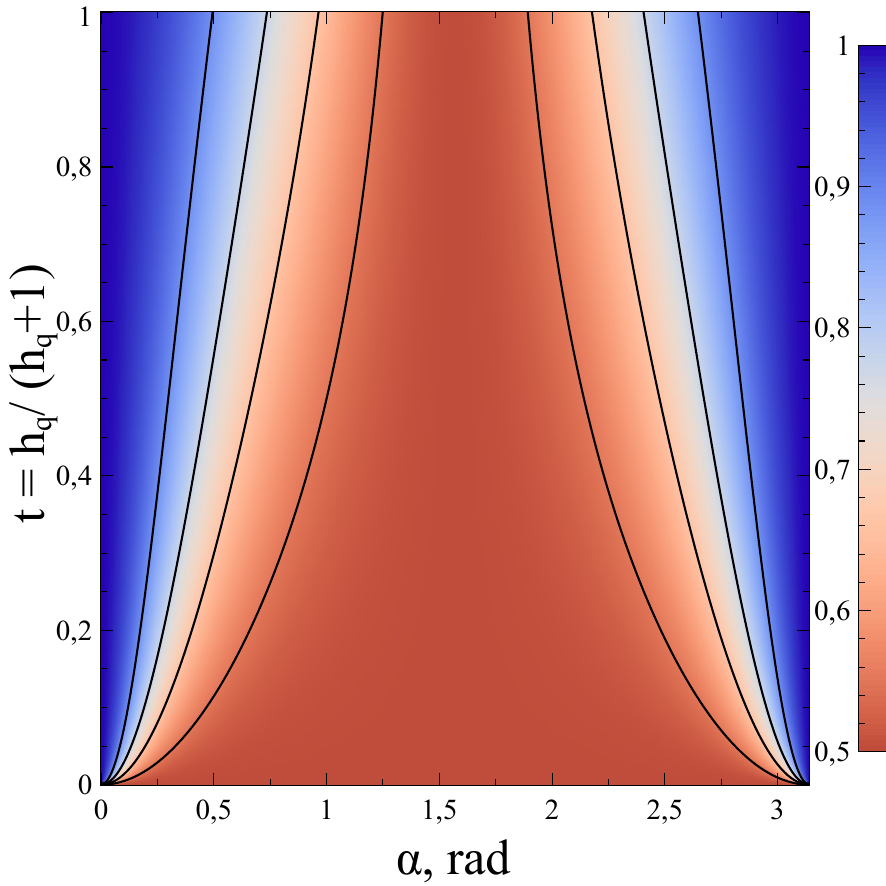}
  \label{fig:std1}
\end{subfigure}\hfill
\begin{subfigure}[t]{0.48\textwidth}
  \centering
  \includegraphics[width=\linewidth]{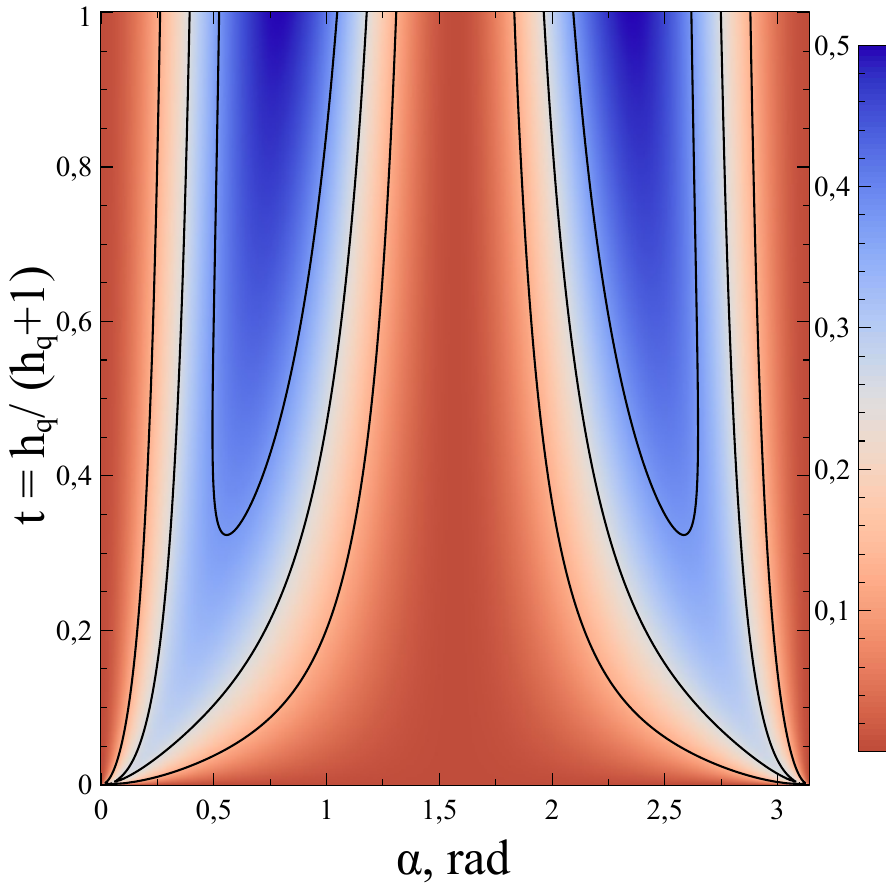}
  \label{fig:std2}
\end{subfigure}

\vspace{6pt}

\begin{subfigure}[t]{0.32\textwidth}
  \centering
  \includegraphics[width=\linewidth]{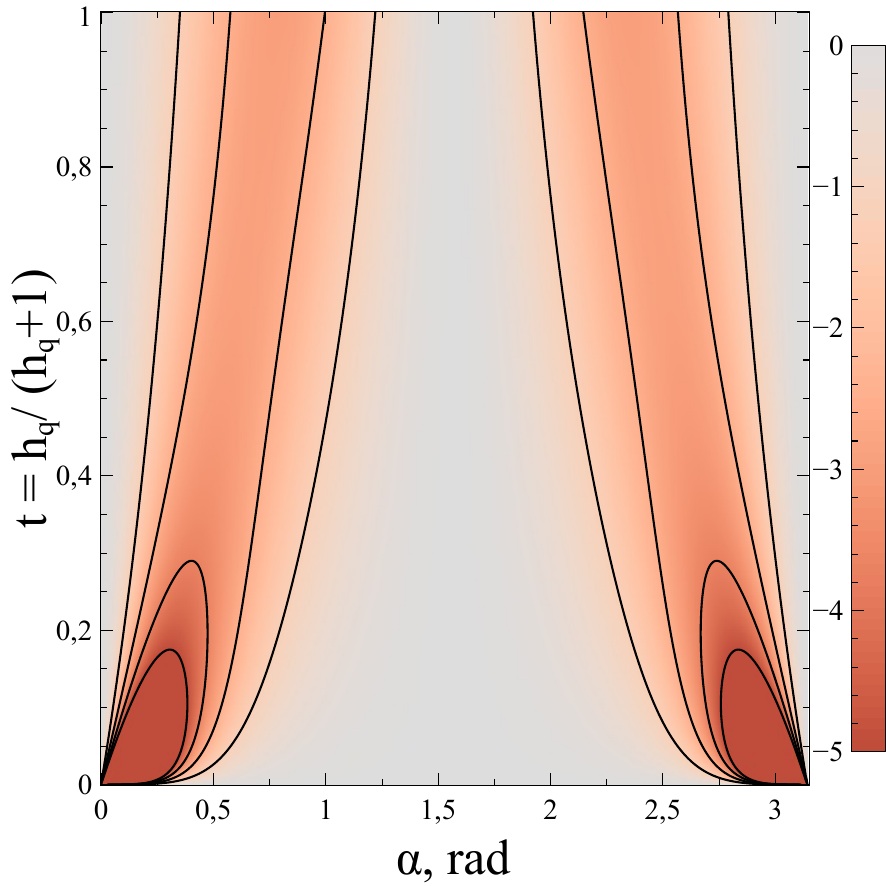}
  \label{fig:hex_xi}
\end{subfigure}\hfill
\begin{subfigure}[t]{0.32\textwidth}
  \centering
  \includegraphics[width=\linewidth]{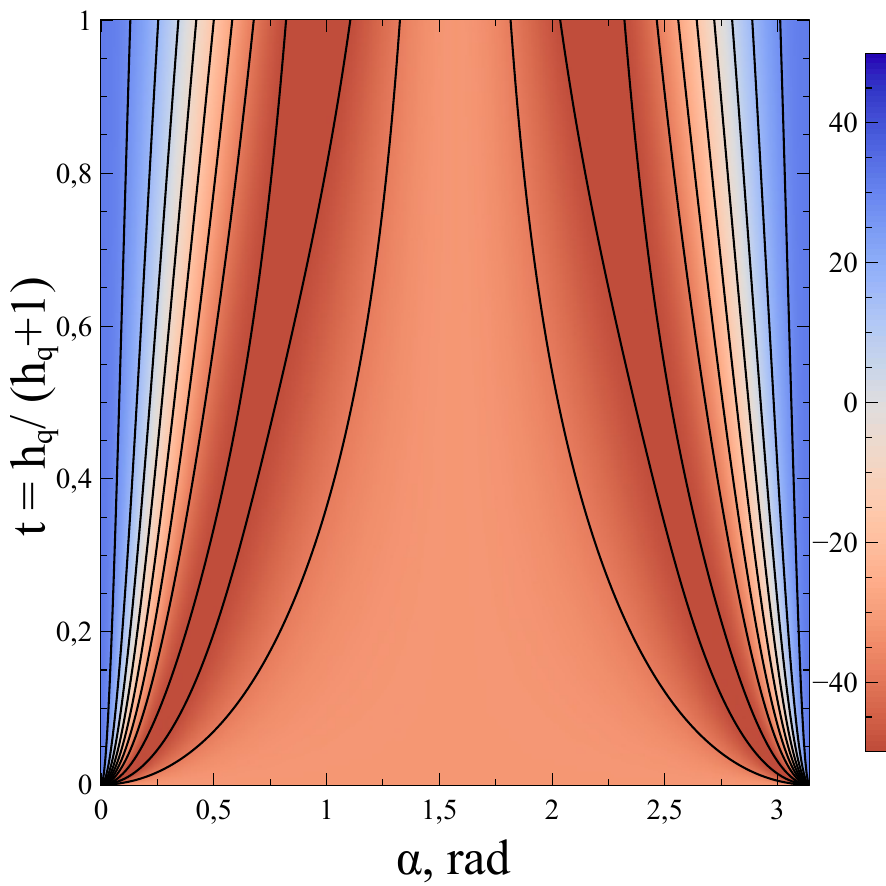}
  \label{fig:hex_xikappa}
\end{subfigure}\hfill
\begin{subfigure}[t]{0.32\textwidth}
  \centering
  \includegraphics[width=\linewidth]{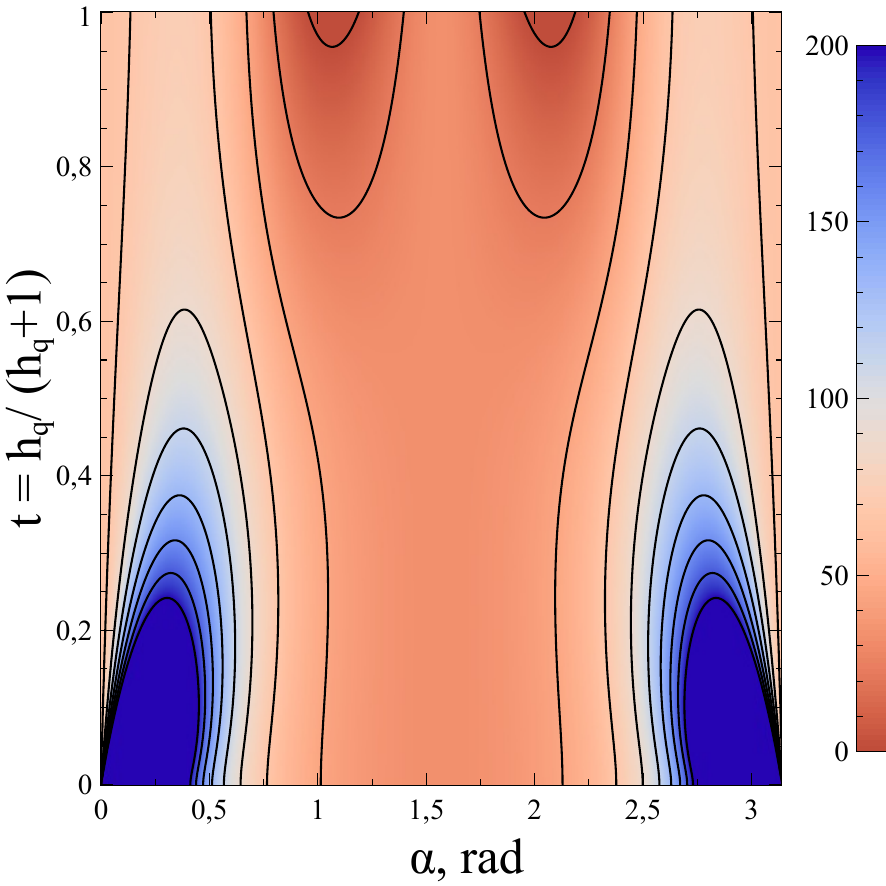}
  \label{fig:hex_xi2}
\end{subfigure}

\caption{Two-dimensional maps in coordinates of $t=h_q/(h_q+1)$ and $\alpha$ of the standard (top row: $h_q^2R_H$ and $h_qR_M$) and hexagonal-exchange-induced (bottom row: $h_qR_{\xi}$, $h_q^2 R_{\xi \kappa}$ and $h_q^2 R_{\xi^2}$) response functions multiplied by a corresponding power of $h_q$ to compensate for the decay at high fields.}
\label{fig:std_and_hex}
\end{figure*}

\subsection*{Response function $R_{\xi}$}

Retaining only the linear terms in the exchange anisotropy parameter $\xi$,
one obtains the response function
\begin{equation}
R_{\xi}
=
-\frac{(h_q + 1)(1 + 3h_q)\sin^{2} 2\alpha}
{(h_q + \sin^{2}\alpha)^{3}},
\end{equation}
with the associated scattering function
$S_\xi = \Lambda_q^{2}\langle \widetilde{I}_M^{2} \rangle q^{2} \xi / 45$, where $\langle \widetilde{I}_M^{2} \rangle$ is the averaged Fourier image of the inhomogeneity function~\cite{MetlovMichels2015}

To visualize the angular and field dependence of $R_{\xi}$, we consider two-dimensional plots in the variables $\alpha$ and $h_q$.
Since $h_q$ varies from zero to infinity, we introduce the transformation
$h_q = t/(1-t)$, which maps the entire field range to the interval
$t \in [0,1)$.
Moreover, because $R_{\xi}$ decays as $1/h_q$ at large fields, we plot the
rescaled quantity $h_q(t) R_{\xi}(h_q(t),\bm q)$.

As shown in Fig.~\ref{fig:std_and_hex}, the response function $R_{\xi}$ exhibits several characteristic features. 
The numerator $\sin^2(2\alpha)$ enforces exact zeros at 
$\alpha=0$, $\pi/2$, and $\pi$, resulting in vertical nodal lines in the detector plane. 
The response reaches its maximum at $\alpha=\pi/4$ and $3\pi/4$, 
leading to two symmetric intensity lobes.

The denominator $(h_q+\sin^2\alpha)^3$ produces a strong enhancement at small $h_q$, 
indicating that the linear symmetric-exchange contribution is primarily a low-field effect. The function remains negative for all physical $h_q$, since $(1+h_q)(1+3h_q)>0$. It is important to emphasize, that although the anisotropic response functions
may assume negative values in certain regions of the $(\alpha,h_q)$ parameter
space, the total magnetic SANS cross section remains strictly positive for all
angles and for all physically relevant values of the model parameters.
This reflects the fact that the anisotropic contributions enter as corrections
to the dominant isotropic background and do not violate the fundamental
positivity of the scattering intensity.

The angular dependence of $R_{\xi}$ is strictly symmetric with respect to the
angles $\pi/2$ and $3\pi/2$.
The response reaches its maximum values at $\alpha = \pi/4$ and $3\pi/4$,
which can be traced back to the dominant high-field angular contribution
$\sin^{2} 2\alpha$.
In addition, $R_{\xi}$ vanishes at $\alpha = 0, \pi/2, \pi$, and $2\pi$.
These zeros become particularly transparent when the response function is
plotted as a one-dimensional function of $\alpha$ at fixed $h_q$.

\subsection*{Response function $R_{\xi\kappa}$}

After performing the directional averaging of the mixed $\xi\kappa$
contribution, the response function in this order can be expressed as a
finite series of even Fourier harmonics,
\begin{equation}
R_{\xi\kappa}
=
\frac{1}{h_q^{2}(h_q + \sin^{2}\alpha)^{3}}
\left(
\chi_{\kappa\xi}^{(0)}
+
\chi_{\kappa\xi}^{(2)} \cos 2\alpha
+
\chi_{\kappa\xi}^{(4)} \cos 4\alpha
+
\chi_{\kappa\xi}^{(6)} \cos 6\alpha
\right),
\end{equation}
with the corresponding scattering function
\begin{equation}
S_{\xi\kappa}
=
\frac{\Lambda_q^{2}}{1260}
\langle \widetilde{I}_M^{2} \rangle
q^{2} \kappa \xi .
\end{equation}
The coefficients $\chi_{\kappa\xi}^{(n)}$ are polynomials of second or third
order in $h_q$, their explicit expressions are listed in Appendix~\ref{app:polynomialschi2}.

The two-dimensional representation of the rescaled responso function
$h_q(t)^{2} R_{\xi\kappa}(h_q(t),\bm q)$, shown in
Fig.~\ref{fig:std_and_hex}, reveals a distinctive feature absent in $R_{\xi}$. 
In contrast to the latter, the mixed response function $R_{\xi\kappa}$ undergoes a sign reversal along a nontrivial angle--field curve
$N(\alpha,h_q)=0$. 
This field-dependent nodal structure originates from the mutual cancellation of several competing angular harmonics and constitutes a characteristic signature of the $\xi\kappa$ contribution. This behavior is qualitatively different from that of the standard response functions
discussed in \cite{Michels2014}.

Although harmonics up to $\cos 6\alpha$ formally enter the numerator, 
the resulting angular variation remains smooth across the detector plane. 
This behavior demonstrates that higher-order harmonics are efficiently suppressed by the denominator and do not generate rapid angular oscillations in the observable scattering pattern.

\subsection*{Response function $R_{\xi^{2}}$}

After averaging the part of the cross section that is quadratic in the exchange anisotropy (given in Appendix~\ref{app:chi2}), the response function $R_{\xi^{2}}$ can likewise be written as a sum of even angular harmonics.
In this case, the expansion also ends up at the sixth harmonic 
\begin{equation}
R_{\xi^{2}}
=
\frac{1}{h_q^{2}(h_q + \sin^{2}\alpha)^{4}}
\left(
\chi_{\xi^{2}}^{(0)}
+
\chi_{\xi^{2}}^{(2)} \cos 2\alpha
+
\chi_{\xi^{2}}^{(4)} \cos 4\alpha
+
\chi_{\xi^{2}}^{(6)} \cos 6\alpha
\right),
\end{equation}
with the corresponding scattering function 
\begin{equation}
    S_{\xi\kappa}
=
\frac{\Lambda_q^{4}}{7560}
\langle \widetilde{I}_M^{2} \rangle
q^{4} \xi^2.
\end{equation}
The explicit expressions for the coefficients $\chi_{\xi^{2}}^{(n)}$ are given
in Appendix~\ref{app:polynomialschi2}.

Owing to the prefactor $h_q^{-2}$, the response is strongly enhanced
in the low-field regime and remains positive across all values in $(t,\alpha)$ parameter space. As $h_q \to 0$, the intensity increases rapidly,
leading to pronounced maxima at small values of the reduced field parameter.
The intensity is concentrated near $\alpha \approx 0$ and $\alpha \approx \pi$,
where $\sin\alpha$ is small and the denominator
$(h_q + \sin^2\alpha)^4$ attains its minimal values.
In contrast, the response is strongly suppressed around
$\alpha = \pi/2$, where $\sin^2\alpha = 1$ and the denominator is maximal.
Unlike the linear response $R_{\xi}$, the function $R_{\xi^2}$ does not
possess a global angular line at which it vanishes for all $h_q$.
Instead, its zeros form curved nodal lines in the $(\alpha,h_q)$ plane,
arising from the mutual cancellation of several even angular harmonics.

As in the case of $R_{\xi \kappa}$, expansion formally contains harmonics up to
$\cos 6\alpha$ but higher-order oscillations are efficiently suppressed and the overall angular dependence remains smooth.

\begin{figure}[h!]
\centering

\begin{subfigure}{0.48\linewidth}
\centering
\includegraphics[width=\linewidth]{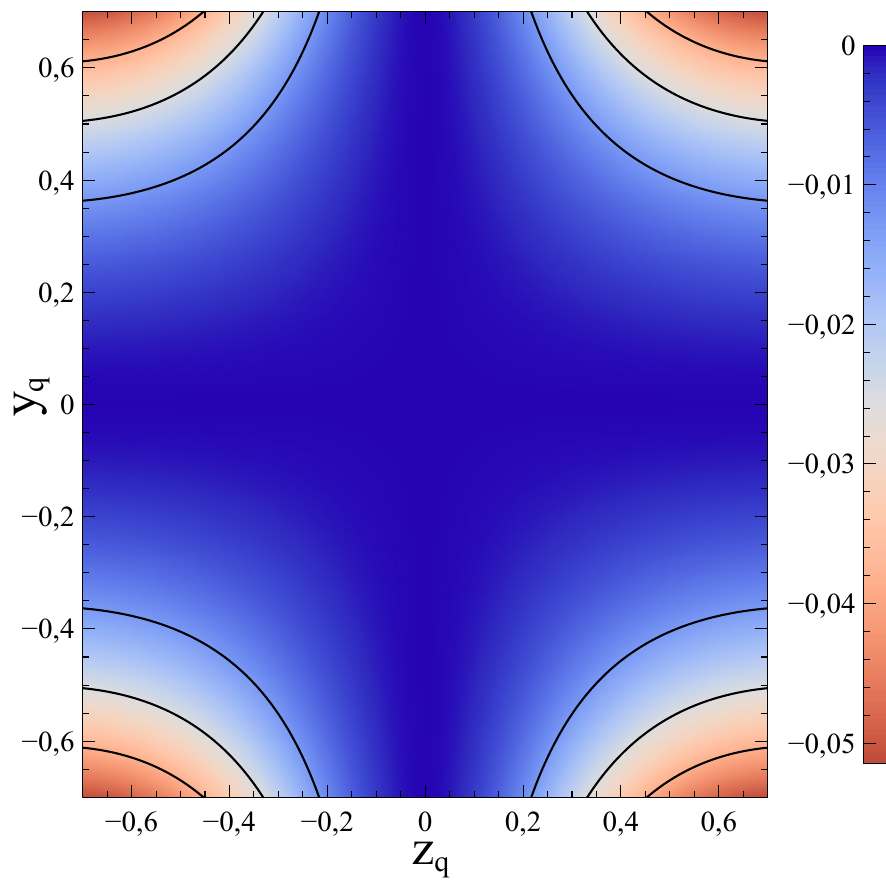}
\caption{Response function $R_{\mathrm{H}}$ at $t=0.5$.}
\end{subfigure}
\hfill
\begin{subfigure}{0.48\linewidth}
\centering
\includegraphics[width=\linewidth]{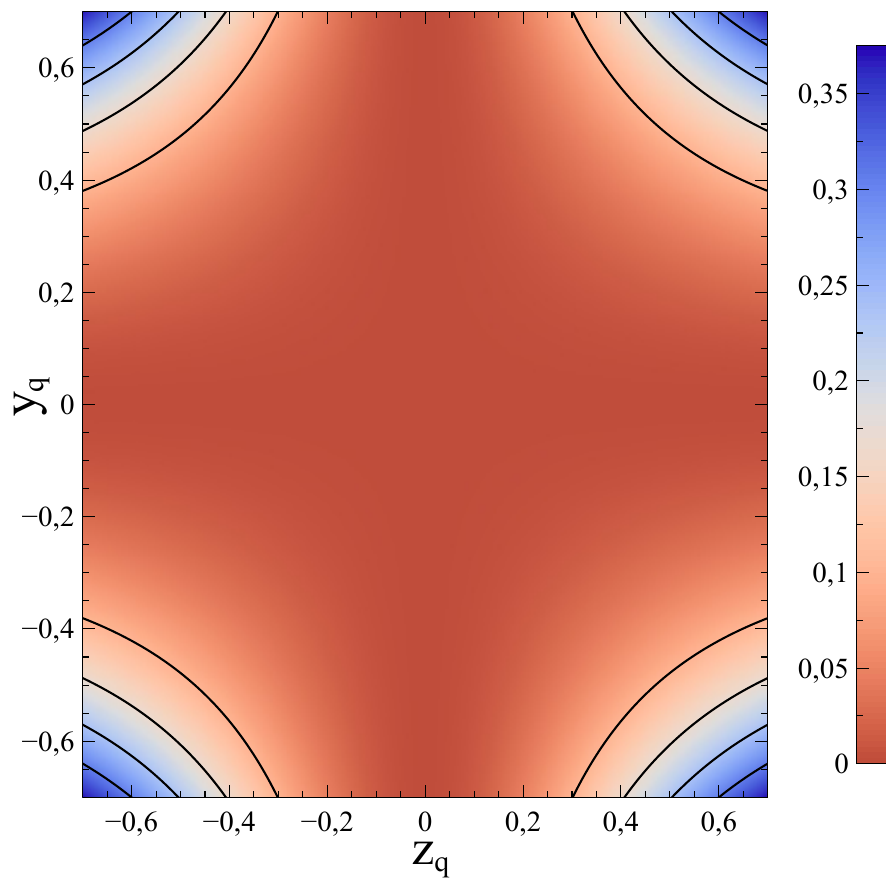}
\caption{Response function $R_{\mathrm{M}}$ at $t=0.5$.}
\end{subfigure}

\caption{
Two-dimensional maps of the standard (isotropic Heisenberg-type) response functions 
in $(y_q,z_q)$ coordinates at the fixed reduced field parameter 
$t=h_q/(h_q+1)=0.5$. 
These plots serve as reference patterns for comparison with the anisotropic 
response functions shown in Fig.~\ref{fig:anisotropic_maps}.
}
\label{fig:standard_maps}
\end{figure}

To facilitate direct comparison with experimental SANS data,
we represent the response functions in detector-plane coordinates
$(z_q,y_q)$ rather than in angular variables.
In typical SANS measurements, the intensity is recorded as a function
of the Cartesian components of the scattering vector in the detector plane.
The $(z_q,y_q)$ representation therefore provides a more transparent
visualization of the angular structure of the response functions
from an experimental perspective.

The standard Heisenberg-type response functions $R_H$ and $R_M$
(as shown on fig.\eqref{fig:standard_maps}) exhibit the characteristic angular anisotropy of
magnetic SANS in the perpendicular geometry.

\begin{figure*}[h!]
\centering

\begin{subfigure}{0.32\linewidth}
\includegraphics[width=\linewidth]{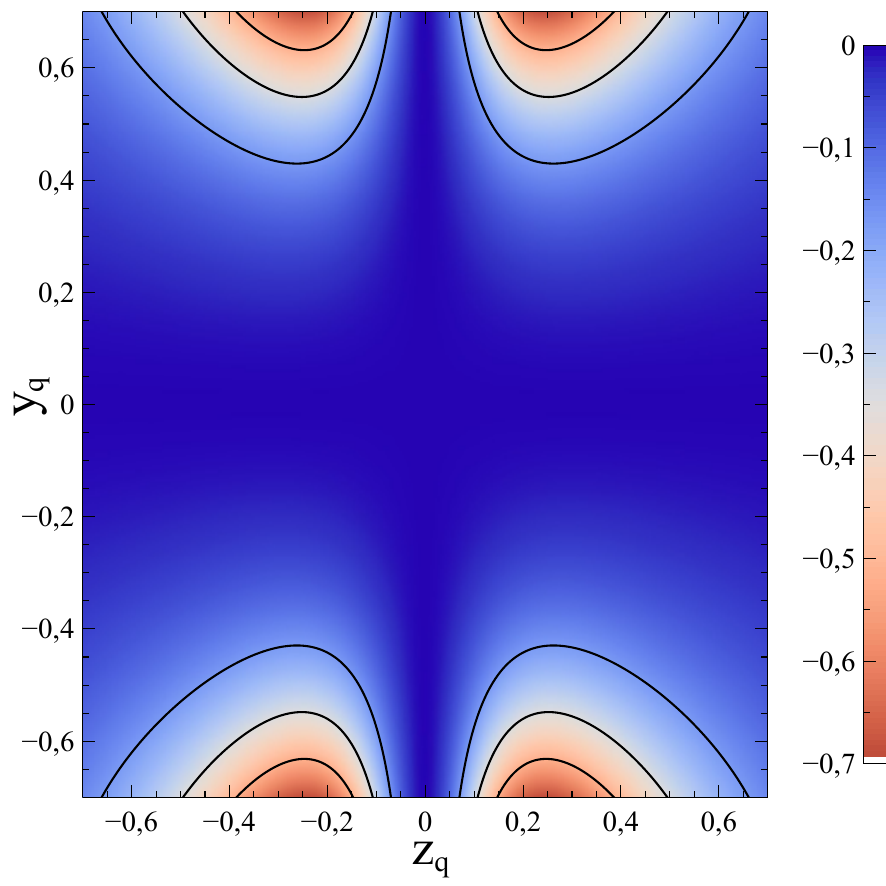}
\end{subfigure}
\hfill
\begin{subfigure}{0.32\linewidth}
\includegraphics[width=\linewidth]{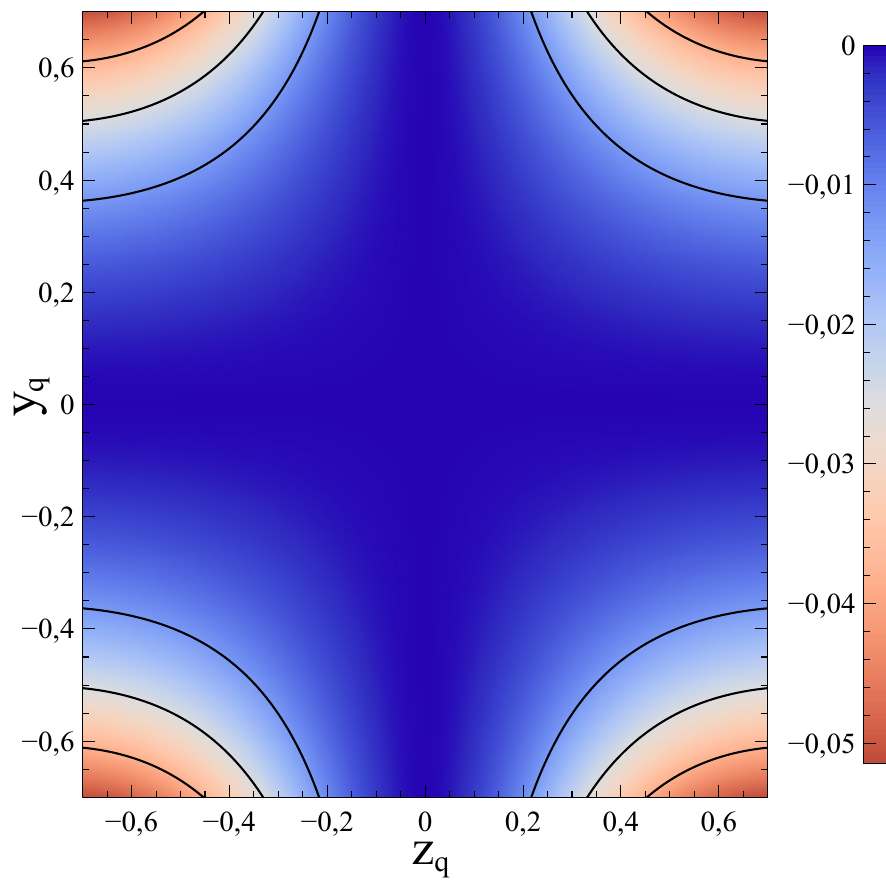}
\end{subfigure}
\hfill
\begin{subfigure}{0.32\linewidth}
\includegraphics[width=\linewidth]{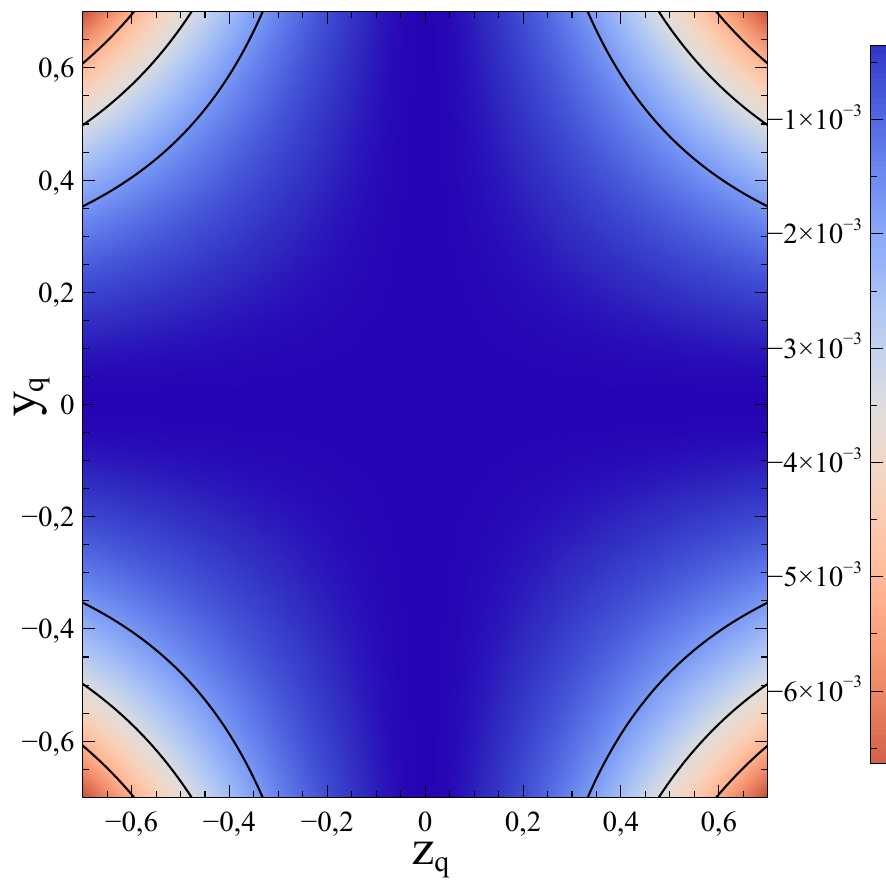}
\end{subfigure}

\vspace{0.3cm}

\begin{subfigure}{0.32\linewidth}
\includegraphics[width=\linewidth]{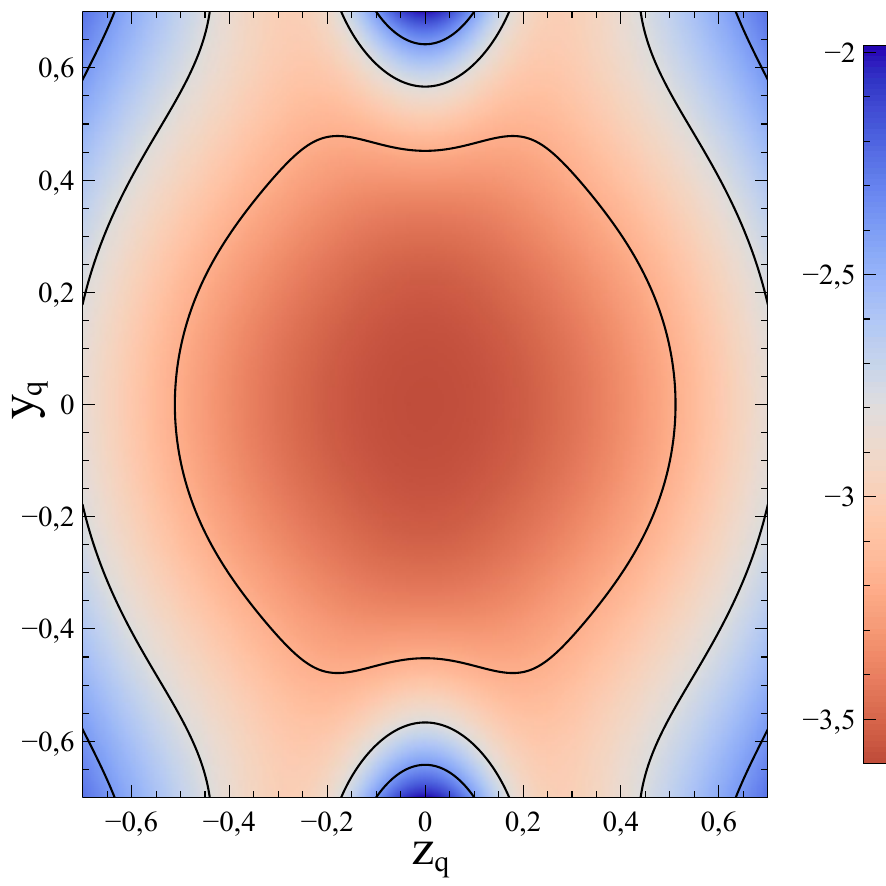}
\end{subfigure}
\hfill
\begin{subfigure}{0.32\linewidth}
\includegraphics[width=\linewidth]{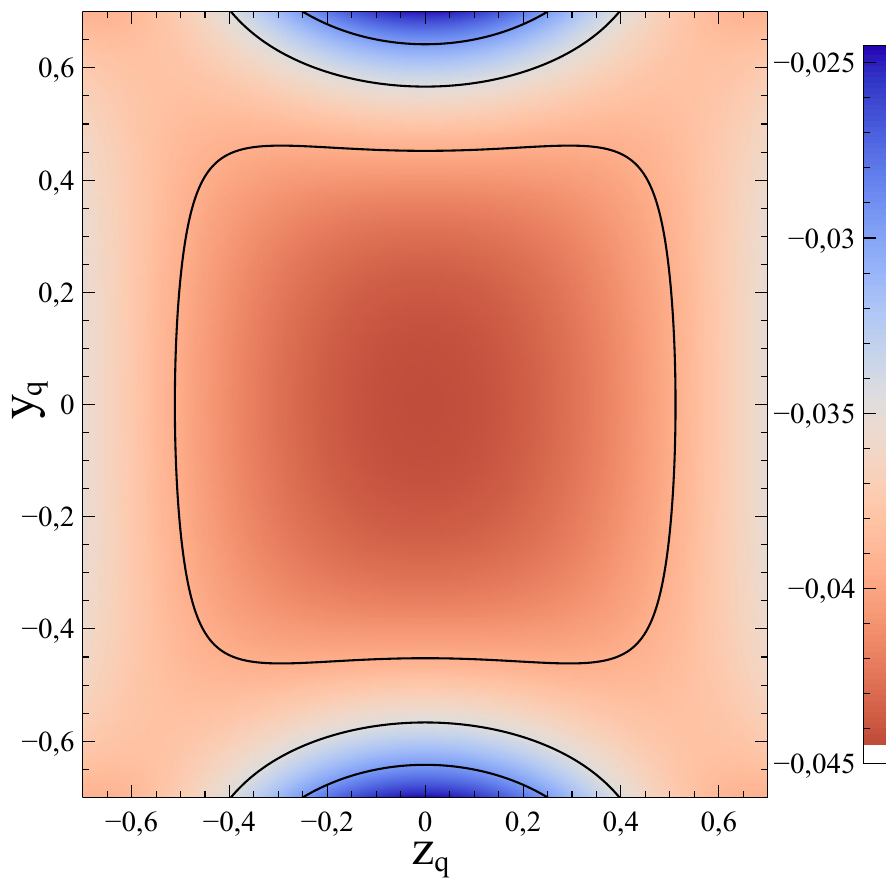}
\end{subfigure}
\hfill
\begin{subfigure}{0.32\linewidth}
\includegraphics[width=\linewidth]{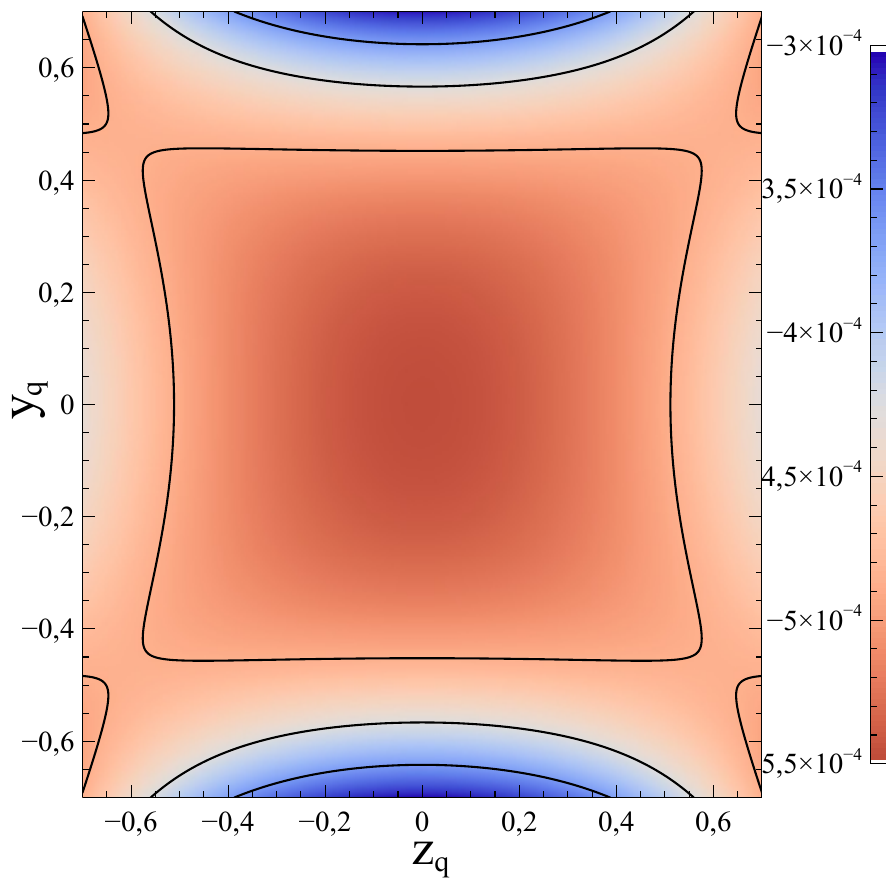}
\end{subfigure}

\vspace{0.3cm}

\begin{subfigure}{0.32\linewidth}
\includegraphics[width=\linewidth]{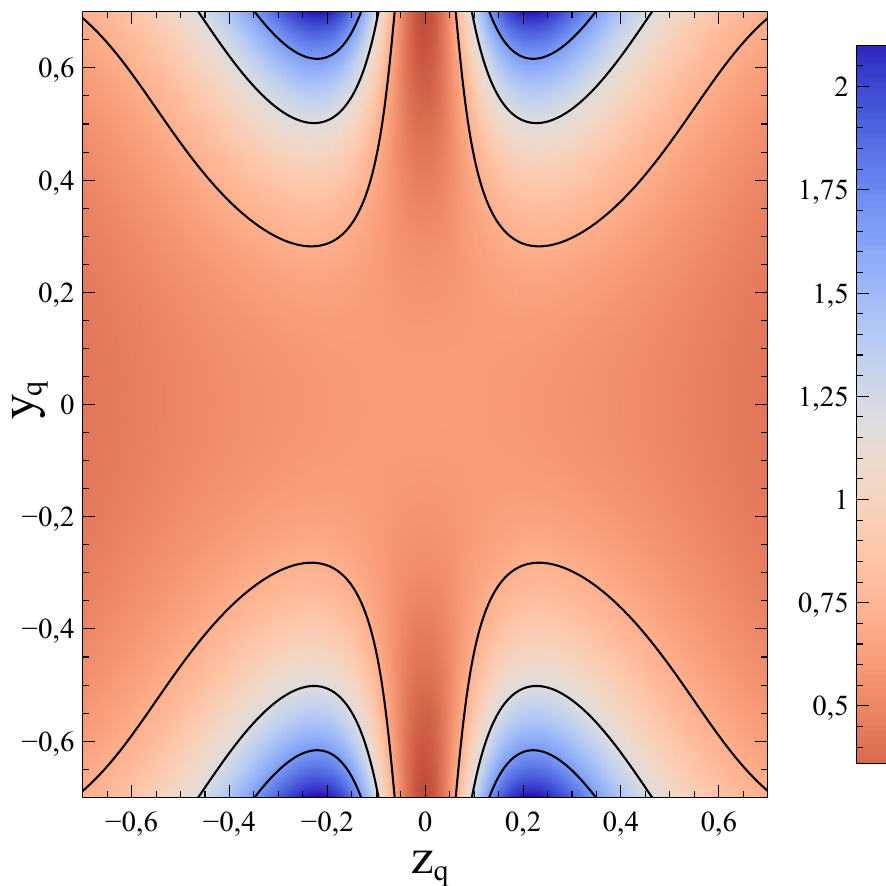}
\caption{$t=0.1$}
\end{subfigure}
\hfill
\begin{subfigure}{0.32\linewidth}
\includegraphics[width=\linewidth]{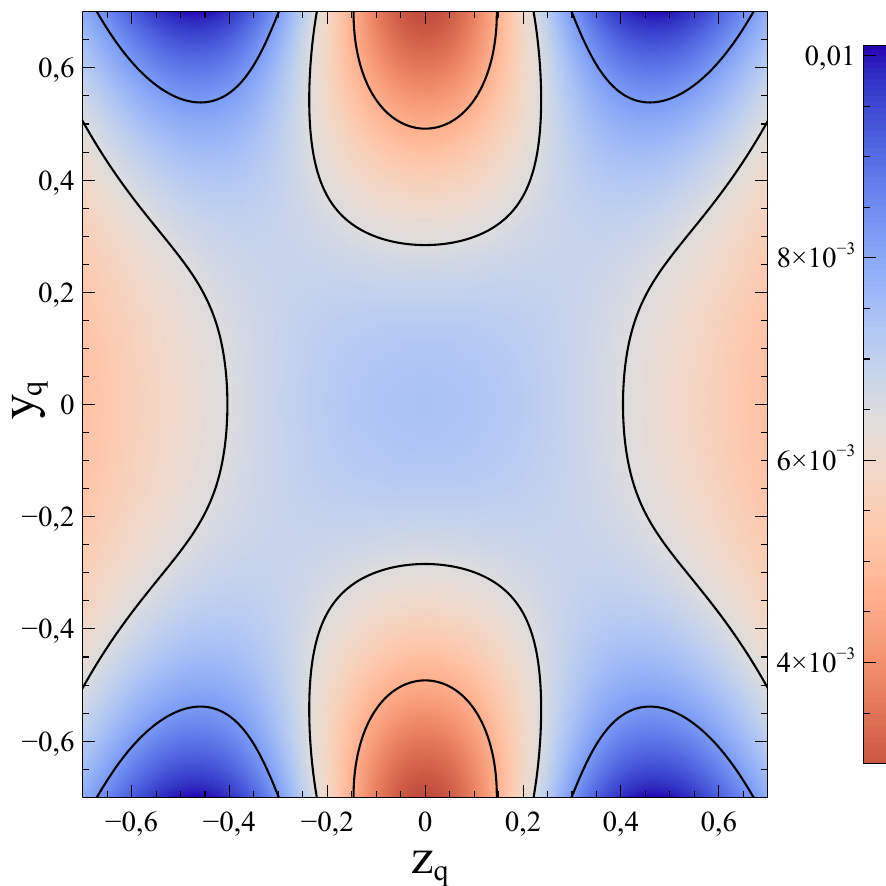}
\caption{$t=0.5$}
\end{subfigure}
\hfill
\begin{subfigure}{0.32\linewidth}
\includegraphics[width=\linewidth]{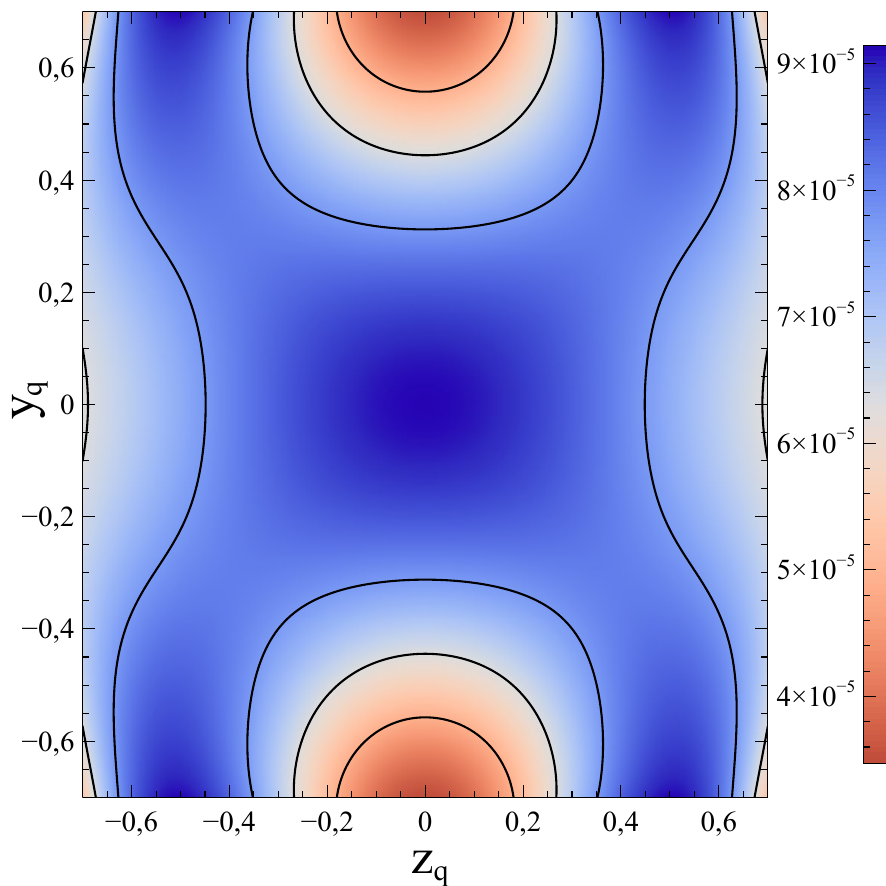}
\caption{$t=0.9$}
\end{subfigure}

\caption{
Two-dimensional maps of the anisotropic response functions in $(y_q,z_q)$ 
coordinates at different values of the reduced field parameter 
$t=h_q/(h_q+1)$. 
Top row: linear contribution $R_{\xi}$. 
Middle row: mixed contribution $R_{\xi\kappa}$. 
Bottom row: quadratic contribution $R_{\xi^2}$. 
Columns correspond to $t=0.1$, $t=0.5$, and $t=0.9$, respectively. 
The evolution of the angular structure with increasing field illustrates 
the distinct symmetry properties and scaling behavior of the individual 
anisotropic terms.
}
\label{fig:anisotropic_maps}
\end{figure*}

In contrast, the exchange-anisotropy-induced terms (shown on fig. \eqref{fig:anisotropic_maps})
display characteristic modifications of this pattern.

The detector-plane maps of $R_M$ and $R_{\xi}$
exhibit qualitatively similar angular structures,
although their signs is opposite.
While  $R_M$ reflects longitudinal
magnetization fluctuations in the Heisenberg-like theory,
the  $R_{\xi}$ arises from the liniar anisotropic
correction of the exchange.
Consequently, linear correction of the symmetric exchange anisotropy does not
introduce a fundamentally new angular symmetry,
but rather redistributes the weight of angular harmonics
already present in the Heisenberg-like response.

The mixed term $R_{\xi\kappa}$, represents the coupling between
symmetric exchange anisotropy and random anisotropy fluctuations.
It vanishes if either $\xi$ or $\kappa$ is zero and therefore
constitutes a genuine interference term between two distinct
sources of magnetic disorder.

The quadratic contribution $R_{\xi^2}$ generates additional angular
modulation. The intensity maxima are aligned predominantly along the
vertical direction ($z_q \approx 0$, $|y_q|$ large),
while suppression occurs near the horizontal axis.

These features constitute experimentally accessible signatures
of symmetric exchange anisotropy.

\subsection{Anisotropic SANS cross sections in parallel geometry}

Substituting equations~\eqref{eq:mX1} and~\eqref{eq:mY1} into the expression for the magnetic SANS
cross section in the parallel geometry and expanding in the small parameter
$\xi$, the cross section again separates into Heisenberg-like and anisotropic
contributions.
In contrast to the perpendicular geometry, however, the contribution linear
in the symmetric exchange anisotropy vanishes identically,
\begin{equation}
\left. \frac{d\Sigma^{M}_{\parallel}}{d\Omega} \right|_{\xi} = 0 .
\end{equation}

The leading anisotropic correction in parallel geometry is therefore given
by the mixed term proportional to $\xi\kappa$.
As in the perpendicular case, this contribution can be written in a compact
form as a scalar product of two vectors,
$\mathbf u = (u_X,u_Y)$ and
$\boldsymbol{\Phi}^{\parallel} = (\Phi_X^{\parallel},\Phi_Y^{\parallel})$,
where the components of $\boldsymbol{\Phi}^{\parallel}$ are given by
\begin{equation}
\Phi_{X}^{\parallel}
=
-
2 h_q^{-2} y_q \big( K_{YZ} x_q - K_{XZ} y_q \big),
\end{equation}
\begin{equation}
\Phi_{Y}^{\parallel}
=
2 h_q^{-2} x_q \big( K_{YZ} x_q - K_{XZ} y_q \big).
\end{equation}

After performing the directional averaging over crystallite orientations,
the corresponding contribution to the magnetic SANS cross section becomes
isotropic in the detector plane and assumes the simple form
\begin{equation}
\left\langle
\frac{d\Sigma^{M}_{\parallel}}{d\Omega}
\right\rangle_{\xi\kappa}
=
-
\frac{8 \Lambda_q^{2} q^{2}}{315 h_q^{2}}
\left\langle \widetilde{I}_M^{2} \right\rangle
\, \xi \kappa .
\end{equation}

The contribution proportional to $\xi^{2}$ is also isotropic.
This follows from the absence of any preferred in-plane direction that could
break the macroscopic rotational symmetry of the system in the detector plane
for the parallel scattering geometry.

\section*{Discussion and Conclusions}

The developed micromagnetic theory of magnetic small-angle neutron
scattering consistently accounts for weak symmetric anisotropic exchange
in centrosymmetric ferromagnets. The exchange interaction was formulated in
terms of a general fourth-rank tensor that was subsequently separated
into isotropic and deviatoric parts. This decomposition makes it possible to
relate all anisotropic effects to the symmetry properties of the deviatoric part of the exchange tensor.

A compact expression derived for the exchange energy and corresponding effective field remains
valid for spatially inhomogeneous saturation magnetization. In the linear
response regime, the isotropic part of the exchange tensor reproduces the
well-known Heisenberg-type SANS response, whereas all anisotropic corrections
are governed exclusively by the deviator of the exchange tensor. In polycrystalline systems, these corrections persist even after directional averaging over crystallites with randomly distributed anisotropy axes, whereby the symmetry of the exchange tensor manifests itself in the observable angular dependencies.

For the case of hexagonal exchange symmetry, we have obtained explicit analytical
expressions for the anisotropic response functions proportional to \(\xi\),
\(\xi\kappa\), and \(\xi^{2}\). The corresponding two-dimensional SANS maps
exhibit several characteristic features absent in Heisenberg-type theory,
including sign reversals, nontrivial nodal curves in angle-field space,
and the emergence of higher even angular harmonics. Despite the presence of
terms up to the sixth harmonic in the analytical expressions, the
angular dependence remains smooth, indicating an efficient suppression of
higher-order harmonics by geometrical and magnetostatic factors.

Inspection of the general expressions for the scattering cross sections (presented in a separate file) shows that the specific symmetry of the exchange tensor, or more precisely of its deviatoric part, does not determine the maximal angular harmonic.
Instead, the role of symmetry is to enforce the vanishing of certain harmonics.

The maximal harmonic is governed by the geometry of the scattering cross section and by the particular contribution under consideration.
For example, for the term proportional to \(\xi^{2}\), the maximal power of the function \(\Delta K_{ij}\) is equal to two.
Since the kernel \(K_{ij}\) can generate at most a second-order polynomial in the components of the scattering vector \(\mathbf q\), the quantity \((\Delta K_{ij})^{2}\) produces terms of order \(q^{4}\).
As a result, the maximal power of \(\mathbf q\) in the expression for the \(\xi^{2}\) contribution to the cross section is eight, once the geometric prefactors originating from magnetostatics are taken into account. Consequently, the total maximal angular order of the cross section is twelve for any symmetry.

The presented formalism is readily extendable to other crystallographic
symmetries. Obtained results establish a link
between symmetric exchange anisotropy and experimentally observable
features in magnetic SANS, offering a route to identify and quantify
anisotropic exchange interactions in nanostructured magnetic materials.

\appendix 
\section{Exchange energy for spatially varying saturation magnetization}
\label{app:derivative}

In the presence of spatially inhomogeneous saturation magnetization, the
normalized magnetization vector entering the exchange energy density~\eqref{eq:ex_energy_density} must be replaced by $\mathbf m = \mathbf M / M_S$.
Accordingly, the spatial derivative of the magnetization component $m_\alpha$
can be written as
\begin{equation}
\partial_\beta m_\alpha
=
\partial_\beta
\left(
\frac{M_\alpha}{M_S}
\right)
=
\frac{1}{M_S}\,\partial_\beta M_\alpha
-
\frac{M_\alpha}{M_S^{2}}\,
\partial_\beta M_S .
\end{equation}
Noting that
\begin{equation}
\frac{\partial_\beta M_S}{M_S}
=
\partial_\beta
\ln\!\left(\frac{M_S}{M_0}\right),
\end{equation}
where $M_0$ is the average saturation magnetization, and the derivative
$\partial_\beta m_\alpha$ can be rewritten in the form
\begin{equation}
\partial_\beta m_\alpha
=
\frac{1}{M_S}
\left[
\partial_\beta M_\alpha
-
M_\alpha
\partial_\beta
\ln\!\left(\frac{M_S}{M_0}\right)
\right].
\end{equation}
This motivates the introduction of the modified derivative
\begin{equation}
\mathcal D_\beta M_\alpha
=
\partial_\beta M_\alpha
-
M_\alpha
\partial_\beta
\ln\!\left(\frac{M_S}{M_0}\right),
\label{eq:modified_derivative_app}
\end{equation}
which allows one to express spatial derivatives of the normalized
magnetization in a compact form,
\begin{equation}
\partial_\beta m_\alpha
=
\frac{1}{M_S}\,\mathcal D_\beta M_\alpha .
\end{equation}
Using this definition, the exchange energy density can be written as
\begin{equation}
e_{\mathrm{EX}}
=
\frac{1}{2 M_S^{2}}
\sum_{\alpha\beta\gamma\delta}
C_{\alpha\beta\gamma\delta}(\mathbf r)\,
\bigl(\mathcal D_\beta M_\alpha\bigr)
\bigl(\mathcal D_\delta M_\gamma\bigr),
\label{eq:exchange_energy_modified_app}
\end{equation}
where $C_{\alpha\beta\gamma\delta}(\mathbf r)$ is the symmetric exchange tensor.
This form of the exchange energy density serves as the starting point
for the derivation of the effective exchange field presented in
Appendix~\ref{app:exchfield}.

\section{Derivation of the effective exchange field}
\label{app:exchfield}

The effective exchange field follows from the functional derivative of the
exchange energy density with respect to the magnetization.
In micromagnetic theory it is defined as~\cite{AharoniBook}
\begin{equation}
H_{\mathrm{EX},\alpha}
=
-\frac{1}{\mu_0}
\left[
\frac{\partial e_{\mathrm{EX}}}{\partial M_\alpha}
-
\partial_\beta
\left(
\frac{\partial e_{\mathrm{EX}}}{\partial(\partial_\beta M_\alpha)}
\right)
\right].
\label{eq:Hex_def_app}
\end{equation}

Fluctuations of the exchange stiffness contribute only to higher-order
corrections in the magnetic SANS cross section~\cite{MetlovMichels2015}.
Therefore, in the following derivation the exchange tensor is treated as a
constant.
We first evaluate the derivative with respect to $\partial_\beta M_\alpha$.
According to the Appendix~\ref{app:derivative}, exchange energy density can be written in the form
\begin{equation}
e_{\mathrm{EX}}
=
\frac{1}{2 M_S^2}
C_{\mu\nu\gamma\delta}
\,
(\mathcal D_\mu M_\nu)
(\mathcal D_\gamma M_\delta),
\end{equation}
which yields
\begin{equation}
\frac{\partial e_{\mathrm{EX}}}{\partial(\partial_\beta M_\alpha)}
=
\frac{1}{2 M_S^2}
C_{\mu\nu\gamma\delta}
\left[
\frac{\partial (\mathcal D_\mu M_\nu)}{\partial(\partial_\beta M_\alpha)}
(\mathcal D_\gamma M_\delta)
+
\frac{\partial (\mathcal D_\gamma M_\delta)}{\partial(\partial_\beta M_\alpha)}
(\mathcal D_\mu M_\nu)
\right].
\label{eq:dedgradM}
\end{equation}
The operator $\mathcal D_\mu M_\nu$ depends on $\partial_\beta M_\alpha$ only
when $\mu=\beta$ and $\nu=\alpha$, since
\begin{equation}
\mathcal D_\beta M_\alpha
=
\partial_\beta M_\alpha
-
M_\alpha \partial_\beta \ln\!\left(\frac{M_S}{M_0}\right).
\end{equation}
It therefore follows that
\begin{equation}
\frac{\partial (\mathcal D_\mu M_\nu)}{\partial(\partial_\beta M_\alpha)}
=
\delta_{\mu\beta}\,\delta_{\nu\alpha}.
\end{equation}
Substituting this result into Eq.~\eqref{eq:dedgradM}, we obtain
\begin{equation}
\frac{\partial e_{\mathrm{EX}}}{\partial(\partial_\beta M_\alpha)}
=
\frac{1}{2 M_S^2}
C_{\alpha\beta\gamma\delta}
(\mathcal D_\gamma M_\delta)
+
\frac{1}{2 M_S^2}
C_{\mu\nu\alpha\beta}
(\mathcal D_\mu M_\nu).
\end{equation}
Upon relabeling dummy indices and using the symmetry of the exchange tensor
$C_{\alpha\beta\gamma\delta}
=
C_{\gamma\delta\alpha\beta}$, this expression can be written in the compact form
\begin{equation}
\frac{\partial e_{\mathrm{EX}}}{\partial(\partial_\beta M_\alpha)}
=
T_{\alpha\beta},
\qquad
T_{\alpha\beta}
=
\frac{C_{\alpha\beta\gamma\delta}}{M_S^2}
\,\mathcal D_\delta M_\gamma .
\label{eq:Aab_def_app}
\end{equation}
The quantity $T_{\alpha\beta}$ plays a role analogous to the stress tensor in
elasticity theory.

We now evaluate the derivative of the exchange energy density with respect to
$M_\alpha$.
Since $\mathcal D_\beta M_\alpha$ contains the term
$-M_\alpha \partial_\beta \ln(M_S/M_0)$, we obtain
\begin{equation}
\frac{\partial (\mathcal D_\beta M_\alpha)}{\partial M_\alpha}
=
-
\partial_\beta
\ln\!\left(\frac{M_S}{M_0}\right).
\end{equation}
Proceeding analogously to the previous steps and again using the symmetry of
the exchange tensor, one finds
\begin{equation}
\frac{\partial e_{\mathrm{EX}}}{\partial M_\alpha}
=
-
\left[
\partial_\beta
\ln\!\left(\frac{M_S}{M_0}\right)
\right]
T_{\alpha\beta}.
\end{equation}
Collecting all contributions and substituting the results into
Eq.~\eqref{eq:Hex_def_app}, the effective exchange field can be written as
\begin{equation}
H_{\mathrm{EX},\alpha}
=
\frac{1}{\mu_0}
\left[
\partial_\beta T_{\alpha\beta}
+
\left(
\partial_\beta
\ln\!\left(\frac{M_S}{M_0}\right)
\right)
T_{\alpha\beta}
\right].
\end{equation}
This expression serves as a starting point in the derivation of the Fourier image of the effective field and subsequent solution of Browns equations.

\section{Expressions for the polynomials $\chi^{(n)}$}
\label{app:polynomialschi}

This appendix collects the explicit expressions for the field-dependent
polynomials $\chi^{(n)}$ that appear in the Fourier harmonic expansions of
the anisotropic response functions $R_{\xi\kappa}$ and $R_{\xi^{2}}$.
These polynomials encode the dependence on the reduced field parameter
$h_q$ and determine the relative weights of the individual angular harmonics.

\subsection*{Polynomials entering the response function $R_{\xi\kappa}$}

The response function $R_{\xi\kappa}$ is expanded in even angular harmonics
up to $\cos 6\alpha$.
The corresponding coefficients are given by
\begin{align}
\chi_{\kappa\xi}^{(0)} &=
-10 - 36 h_q - 70 h_q^{2} - 24 h_q^{3}, \\
\chi_{\kappa\xi}^{(2)} &=
15 + 48 h_q + 45 h_q^{2} + 32 h_q^{3}, \\
\chi_{\kappa\xi}^{(4)} &=
-6 - 12 h_q + 22 h_q^{2} + 24 h_q^{3}, \\
\chi_{\kappa\xi}^{(6)} &=
1 + 3 h_q^{2}.
\end{align}

\subsection*{Polynomials entering the response function $R_{\xi^{2}}$}
\label{app:polynomialschi2}

The response function $R_{\xi^{2}}$ also contains even angular harmonics up to
$\cos 6\alpha$.
The coefficients multiplying these harmonics are given by
\begin{align}
\chi_{\xi^{2}}^{(0)}
&=
10
+
4h_q\!\left(
12
+
h_q
\left(
30
+
h_q (20 + 9 h_q)
\right)
\right),
\\[4pt]
\chi_{\xi^{2}}^{(2)}
&=
-15
+
2h_q\!\left(
-32
+
h_q
\left(
-48
+
h_q (-26 + 17 h_q)
\right)
\right)),
\\[4pt]
\chi_{\xi^{2}}^{(4)}
&=
2\!\left(
3
+
2h_q(-2+hq)
(-2+hq(2+3hq)
)
\right),
\\[4pt]
\chi_{\xi^{2}}^{(6)}
&=
-1
-
6h_q^3
\left(
2
+
3 h_q
\right)
.
\end{align}
The higher angular harmonics present in $R_{\xi^{2}}$ originate from the
additional angular prefactors discussed in Sec.~VIII.

\section{Expression for the $\xi^2$ part of the anisotropic cross section}
\label{app:chi2}

This appendix provides the explicit expression for the contribution to the
perpendicular magnetic SANS cross section that is quadratic in the 
exchange anisotropy parameter $\xi$.
The result is obtained by expanding the general expressions derived in
Sec.~VI to second order in $\xi$ and retaining only terms that contribute
within the linearized micromagnetic approximation

\begin{align}
\left. \frac{d\Sigma^{M}_{\perp}}{d\Omega} \right|_{\xi^2} = \frac{K_{XZ}^{2}}{h_q^{2}}
&-
\frac{
2 y_q
\Big[
h_q^{2} K_{YY} K_{YZ}
+
(1+h_q) K_{XY} K_{XZ} (h_q + y_q^{2})
\Big]
z_q
}{
h_q^{2} (h_q + y_q^{2})^{2}
}
\nonumber\\[4pt]
&+
\frac{
\Big[
h_q^{3} K_{YZ}^{2}
+
\big(
h_q K_{XY}^{2}
+
h_q^{2}
(2K_{XY}^{2}+2K_{YY}^{2}+K_{YZ}^{2})
\big) y_q^{2}
+
(1+2h_q) K_{XY}^{2} y_q^{4}
\Big]
z_q^{2}
}{
h_q^{2} (h_q + y_q^{2})^{3}
}
\nonumber\\[4pt]
&-
\frac{
2 y_q
\Big[
h_q K_{XY} K_{XZ}
+
2 h_q K_{YY} K_{YZ}
+
K_{XY} K_{XZ} y_q^{2}
\Big]
z_q^{3}
}{
h_q (h_q + y_q^{2})^{3}
}
\nonumber\\[4pt]
&+
\frac{
y_q^{2}
\Big[
h_q (2K_{XY}^{2}+3K_{YY}^{2})
+
2 K_{XY}^{2} y_q^{2}
\Big]
z_q^{4}
}{
h_q (h_q + y_q^{2})^{4}
}.
\label{eq:Sigma_xi2}
\end{align}

The expression above serves as the starting point for the directional
averaging performed in Sec.~VII and for the construction of the corresponding
response function $R_{\xi^2}$ discussed in the main text. For specific crystallographic symmetries, many terms in
Eq.~\eqref{eq:Sigma_xi2} vanish identically after directional averaging,
leading to the compact harmonic expansions discussed in Sec.~VIII.

\begin{acknowledgements}
The authors are grateful to Prof. Andreas Michels from the University of Luxembourg for careful reading of the manuscript and many valuable remarks.
\end{acknowledgements}

\ConflictsOfInterest{The authors declare that there are no conflicts of interest.}

\DataAvailability{In this theoretical work all the data (to reproduce figures and other results of the present study) are contained within the manuscript.}


\end{document}